\documentclass[11pt]{article}
\usepackage{amsmath,amsbsy,amsfonts,amssymb,graphics,epsfig,float,mathrsfs,amsthm,braket}
\usepackage{ifsym}
\usepackage{psfrag}
\usepackage{color}
\usepackage{caption}
\usepackage{subcaption}
\usepackage{multirow}
\usepackage{float}
\usepackage[normalem]{ulem}

\newcommand \beq{\begin{equation}}
\newcommand \eeq{\end{equation}}
\newcommand{\Utot}{U_{\mathrm{tot}}}
\newcommand{\Zend}{w_{\mathrm{end}}}
\newcommand{\alphaBar}{\alpha_{\mathrm{Barois}}}
\newcommand{\Lp}{L_p}
\newcommand{\Lpg}{L_p^{g}}
\newcommand{\tLpg}{\tilde{L_p^{g}}}
\newcommand{\kappac}{\kappa^{(c)}}
\newcommand{\kappaLc}{\kappa_L^{(c)}}
\newcommand{\leg}{\ell_{eg}}

\newcommand{\upd}{\mathrm{d}}

\newcommand{\s}[1]{{\textsf{\textbf{#1}}}}

\begin{document}
\title{\s{Limitations of curvature--induced rigidity: How a curved strip buckles under gravity}}
\author{ \textsf{Matteo Taffetani$^{1}$, Finn Box$^{1}$, Arthur Neveu$^{1}$ and Dominic Vella$^{1}$}\\ 
	{\it$^{1}$Mathematical Institute, University of Oxford, UK}}

\date{\today}
\maketitle
\hrule\vskip 6pt
\begin{abstract}
The preference of thin flat sheets to bend rather than stretch, combined with results from Geometry, mean that changes in a thin sheet's Gaussian curvature are prohibitively expensive. As a result, an imposed curvature in one principal direction inhibits bending in the other: so-called curvature-induced rigidity. Here, we study the buckling behaviour of a rectangular strip of finite thickness held horizontally in a gravitational field, but with a transverse curvature  imposed at one end.  The finite thickness of the sheet limits the efficacy of curvature-induced rigidity in two ways: (i) finite bending stiffness acts to `uncurve' the sheet, even if this costs some stretching energy, and (ii) for sufficiently long strips, finite weight deforms the strip downwards, releasing some of its gravitational potential energy. We find the critical imposed curvature required to prevent buckling (or, equivalently, to rigidify the strip), determining the dependence on geometrical and constitutive parameters, as well as describing the buckled shape of the strip well beyond the threshold for buckling. In doing so, we quantify the intuitive understanding of curvature-induced rigidity that we gain from curving the crust of a slice of pizza to prevent it from drooping downwards as we eat.
\end{abstract}
\vskip 6pt
\hrule

\maketitle

\section{\label{SEC:Introduction}Introduction}
\begin{figure}
	\centering
	\includegraphics[width=0.65\linewidth]{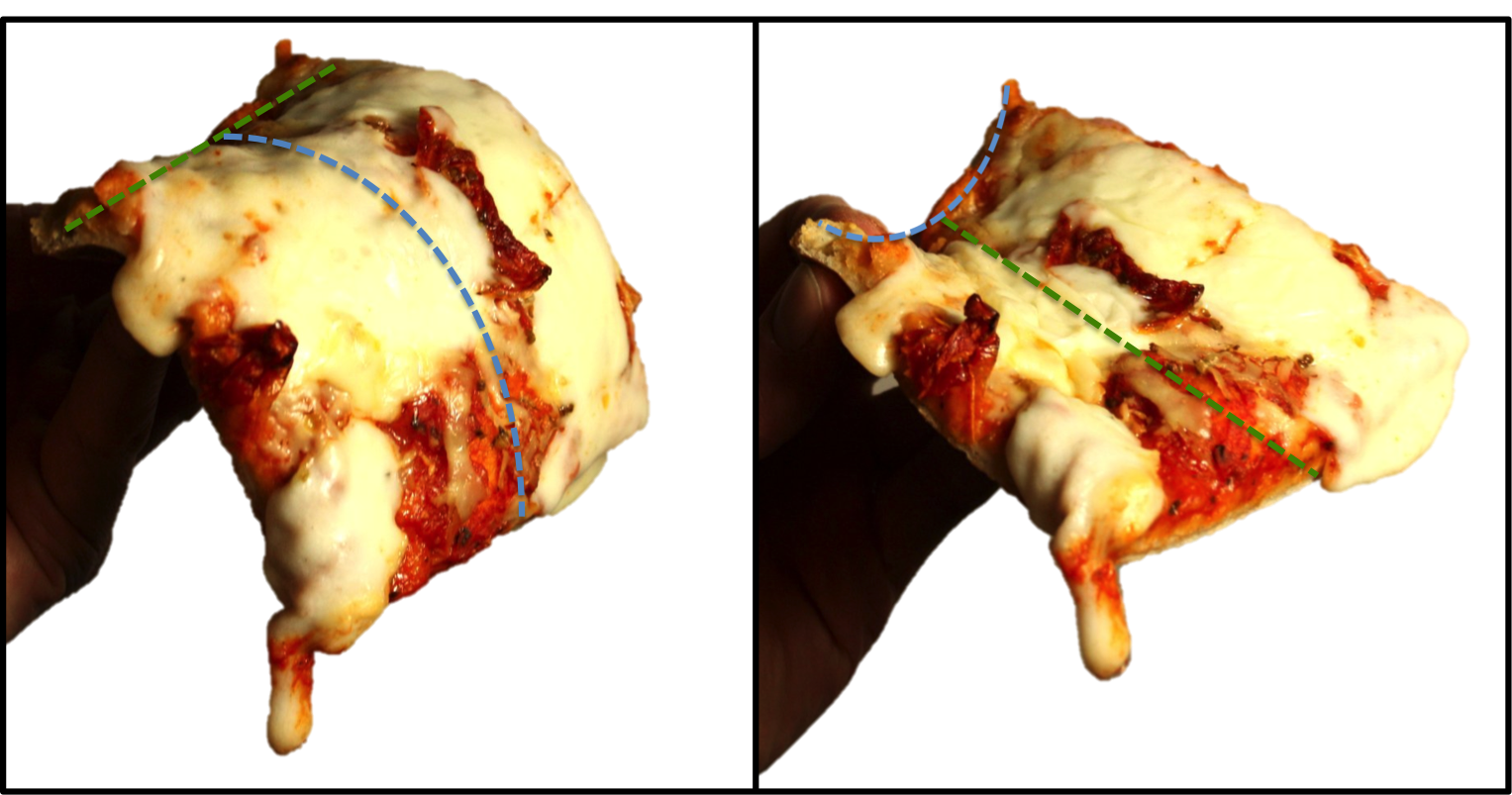}
	\caption{ A slice of pizza droops under its own weight (left). To prevent drooping, one instinctively bends the crust; this imposed curvature increases the effective rigidity of the slice, preventing drooping (right).}\label{FIG:examples}
\end{figure}

From everyday experience, one knows that bending the crust of a slice of pizza prevents it from drooping under its weight. Insight into this intuitive solution to a meal-time conundrum comes from Gauss' \emph{Theorema Egregium} \cite{Wilson2008,Numberphile,Holmes2019}, which states that changes in Gaussian curvature require an energetic cost associated with stretching. In the case of thin sheets, for which bending is energetically favourable over stretching, imposing curvature in the transverse direction induces a resistance to bending in the longitudinal direction since the sheet must then stretch in order to deform (see fig.~\ref{FIG:examples}).

This curvature--induced rigidity \cite{Pini2016} is limited, however, by the finite thickness of the sheet. A naturally flat rectangular strip can be bent isometrically into a cylindrical shape, maintaining zero Gaussian curvature,  only if suitable bending moments are prescribed along all four edges to counter the bending resistance of the strip. This resistance to bending means that, without any imposed edge-moments, a strip with finite thickness will try to `uncurve', even at the expense of incurring some stretching energy. For large imposed curvature, Barois \emph{et al.}~\cite{Barois2014_prl} showed that the  transverse curvature relaxes with distance from the end at which it is imposed, vanishing beyond a persistence length $L_p$ that is determined by a balance between the curvature and stretching of the strip. This persistence length is given in scaling terms as 
\beq
\frac{\Lp}{W}\sim \frac{W}{(tR)^{1/2}}
\label{eqn:BaroisLp}
\eeq
where $W$ and $t$ are the width and thickness of the strip, respectively, while $R$ is the imposed radius of curvature. 

A second consequence of finite thickness is a non-zero weight: a strip held horizontally in a gravitational field also deforms longitudinally (\emph{i.e.}~in the direction orthogonal to the imposed transversal curvature) to release some of its gravitational potential energy. In this paper, we study how these two consequences of finite thickness combine to limit the efficacy of curvature-induced rigidity in a gravitational field.
We anticipate that  as the ratio of bending to gravitational energies decreases, the longitudinal curvature in the strip increases. It is therefore instructive to examine the total energy of the system as a function of the longitudinal curvature. A description of the relations between stability, curvatures and convexity of this energy functional can be found in related studies\cite{Mansfield2017,Giomi2012}. For a clamped transversely-flat cantilever beam \cite{AudolyPomeau}, the total energy functional is a convex function of the imposed curvature: increases in the longitudinal curvature result in increased, yet continuous, bending of the beam. In contrast, a strip with uniform transverse curvature (\emph{e.g.}~a meter tape) can buckle suddenly \cite{Ponomarenko2012}, attaining large changes in shape as a result of small changes in longitudinal curvature. This potential for buckling is a direct consequence of the non-convexity of the total energy functional and corresponds physically to a change in sign of the derivative of the moment required to impose a particular curvature. The situation we consider here has properties that lie between these two limiting cases, and the longitudinal curvature is induced by gravity acting on the strip of finite thickness. It is therefore natural to ask how a strip with imposed curvature at one end will respond to increasing gravitational loading or, equivalently, decreasing the imposed transverse curvature with fixed gravitational loading.

\begin{figure}[ht]
	\centering
	\includegraphics[width=0.65\linewidth]{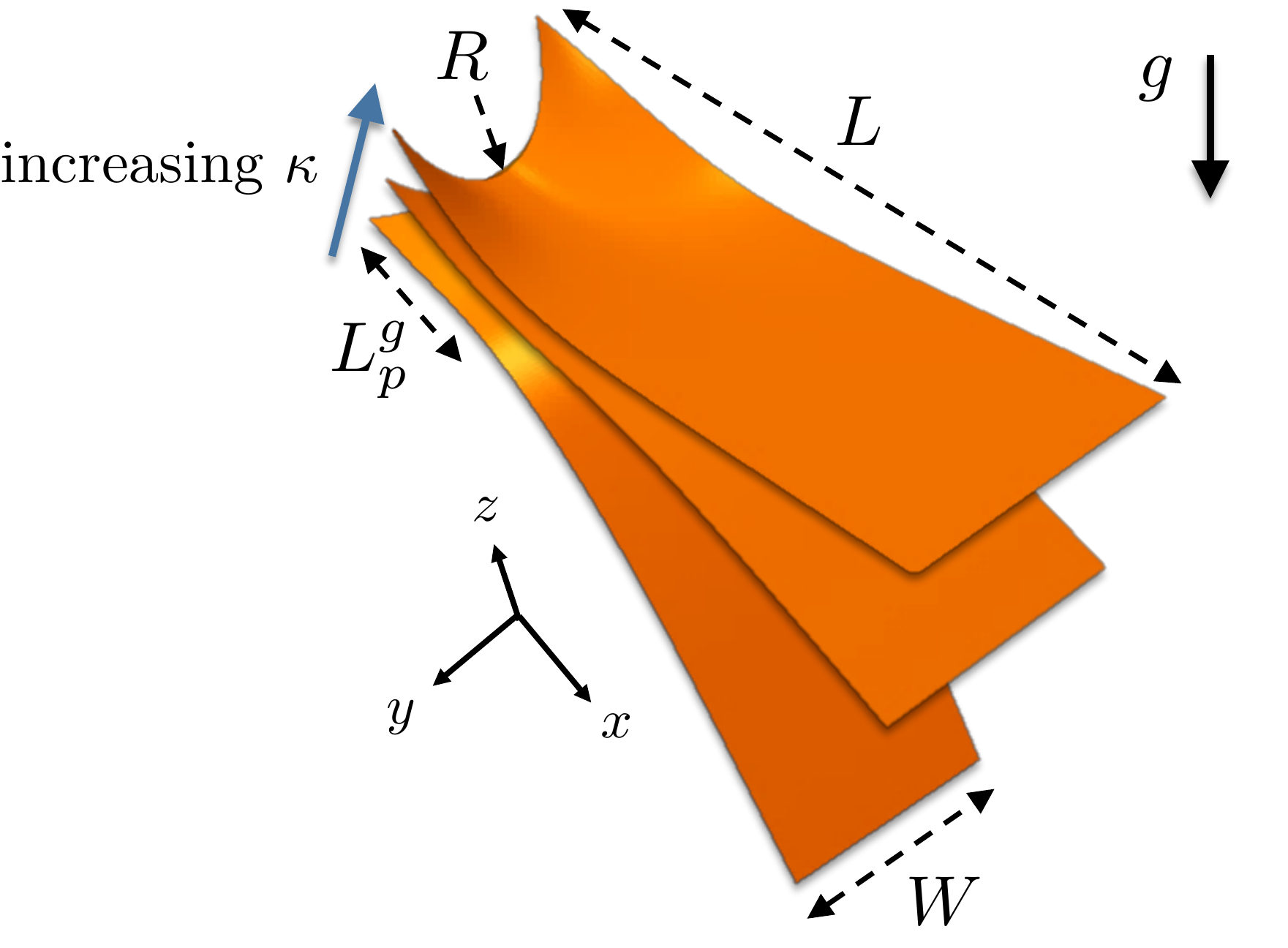}
	\caption{Geometry of an elastic strip and its shape accommodation as a response to an increasing imposed curvature for a given gravitational load. For large enough imposed curvature, the strip is effectively rigidified against its own weight.}
	\label{FIG:geometry}
\end{figure}

The geometry we consider is shown in fig.~\ref{FIG:geometry}. There are four length scales in the problem: the geometrical properties of the strip (its length $L$, width $W$ and thickness $t$), together with the radius of curvature that is imposed at one end, $R$. From the first three lengths, we construct two dimensionless parameters that characterize the geometry of the strip:
\beq
\lambda=\frac{L}{W},\quad \tau=\frac{t}{L}
\eeq To rescale the imposed radius of curvature $R$ we bear eq.~\eqref{eqn:BaroisLp} in mind and let 
\beq
\kappa=\frac{W^2}{8Rt}
\label{eqn:kappa}
\eeq  (the factor $8$ is chosen so that $\kappa={\cal Z}_0/t$ with ${\cal Z}_0=(W/2)^2/(2R)$ the apparent thickness of the curved strip  at the end where curvature is imposed). With this choice, eq.~\eqref{eqn:BaroisLp} becomes $L_p/W\sim\kappa^{1/2}$, and is found to hold provided that $\kappa\gtrsim5$ \cite{Barois2014_prl}.

To account for the role of gravity, we note that our strip has Young's modulus $E$ and density $\rho$, and  denote the gravitational acceleration by $g$. By balancing the bending modulus $B = E t^3 W/12$ of a strip of rectangular cross section  with the weight of the strip per unit  length $\rho g t W$, we extract an elasto-gravitational length $\leg=(B/\rho g t W)^{1/3}$ \cite{Holmes2019,Wang1986}. The relative strength of gravity is then given by the dimensionless parameter
\beq
G= \frac{L^3}{\leg^3} = \frac{\rho g t W}{B} L^3 = \frac{12 \rho g L^3}{E t^2}.
\eeq
(Note that here we have taken the classical choice of the elasto-gravitational length, based on the heavy elastica equation\cite{Wang1986}. However, we shall see that buckling is, in fact, controlled by the value of $G/\tau$, rather than $G$ alone.) 

Given the plethora of dimensionless parameters in this problem, there are numerous ways in which limitations on the effectiveness of the curvature-induced stiffness could be quantified. 
In what follows, we shall assume that the strip's properties are given, so that the question reduces to determining the critical value of the curvature, $\kappa=\kappac(\lambda,\tau,G)$, at which the strip is effectively rigidified.

\section{\label{SEC:BucklingBending} A simple view of the buckling threshold}

The state of the system is determined by a balance between a number of energetic terms. To better understand the balances involved we first recall the derivation \cite{Barois2014_prl} of the persistence length  in the absence of gravity, $\Lp^0$: this is obtained at a scaling level  by optimizing the sum of the stretching energy $U_S\sim Et\cdot ({\cal Z}_0^2/\Lp^2)^2\cdot\Lp W$ (which penalizes small $\Lp$) and the bending energy $U_B\sim B\cdot(1/R)^2\cdot \Lp$  (which penalizes large $\Lp$). Minimizing the sum $U(G=0)=U_B+U_S$ leads directly to the scaling  \eqref{eqn:BaroisLp}, while a more detailed calculation \cite{Barois2014_prl} suggests that the prefactor in eq. \eqref{eqn:BaroisLp} is $\alphaBar=2/\sqrt{35}$. In the presence of gravity, however, we expect that the gravitational potential energy of the strip will play an important role,  possibly reducing the length of the curved region, $\Lpg$, from its value in the absence of gravity. As a simple `toy' model of the effect of gravity in this problem, we consider the gravitational potential energy of just the flat portion of the strip (beyond the part that is curved), \emph{i.e.}~$\Lpg<x<L$. Modelling this portion as an uncurved cantilever, the vertical deflection of the strip's end under gravity is  $\Zend\sim \rho g t W(L-\Lpg)^4/B$; the gravitational potential energy of this portion of the strip is thus $U_G\sim(\rho g t W)^2(L-\Lpg)^5/B$. Combining these  energies, we have the total energy $U_{\mathrm{tot}}$, which may be written in dimensionless form as
\beq
\label{eqn:scalingToy}
\frac{U_{\mathrm{tot}}}{Et^5w /L^3}=-\alpha_G\left(\frac{G}{\tau}\right)^2(1-\tLpg)^5+\alpha_S\frac{\kappa^4}{(\tLpg)^{3}}+\alpha_B\lambda^4\kappa^2\tLpg
\eeq where $\tLpg=\Lpg/L<1$ is the proportion of the strip that remains curved. (We have explicitly introduced the constants of proportionality $\alpha_G$, $\alpha_S$ and $\alpha_B$ to label the relevant terms as originating with the gravitational, stretching and bending energies, respectively.)

To understand the effect of gravity better, we assume that its effect is perturbative, that is that $\tLpg=\tilde{\Lp^0}+\tilde{\delta L}$ for some $\tilde{\delta L}/\tilde{\Lp^0}\ll1$. Minimizing the energy in eq.~\eqref{eqn:scalingToy} we find that $\Lp^0=(3\alpha_S/\alpha_B)^{1/4}\kappa^{1/2}W$, in accord with the scaling of \eqref{eqn:BaroisLp} and, further, that
\beq
\tLpg-\tilde{\Lp^0}\sim-\frac{G^2}{\tau^2}\frac{(1-\tilde{\Lp^0})^4({\tilde{\Lp^0}})^5}{\kappa^4}
\eeq
As expected, this leading order expression shows that the effect of gravity is to reduce the portion of the strip that remains curved, decreasing $\Lpg$ from its value in the absence of gravity. To see the effect of this decrease in $\Lp$, note that the vertical deflection of the strip's end, $ \Zend\sim -\rho g t W (L-\Lpg)^4/B$, and hence 
\beq
\Zend\sim - G\frac{(L-\Lp^0)^4}{L^3}\left[1+\alpha\frac{G^2}{\tau^2\lambda^8}\left(\frac{L-\Lp^0}{\Lp^0}\right)^3\right]
\eeq for some constant $\alpha$.
Hence, as the strength of the gravitational field increases, the deflection of the far end increases also, as should be expected. To highlight the effect of the gravity-induced lengthening of the flat portion (shortening of the curved region), we consider the rate at which $|\Zend(G)|$ increases as $G$ increases: even for small $G$, increases in $G$ increase $|\Zend|$ since the flat end is displaced. However, the shortening of the curved region enhances this effect, as can be seen by considering the rate of change of $\Zend(G)$, $\Zend'(G)$, relative to its value as $G\to0$. We find that
\beq
\frac{\Zend'(G)}{\Zend'(0)}-1\sim \frac{G^2}{\tau^2\lambda^8}\left(\frac{L-\Lp^0}{\Lp^0}\right)^3.
\eeq In particular,  when $G=G_c$ with
\beq\label{EQ:scaling}
\frac{G_c}{\tau\lambda^4}\sim \left(\frac{L-\Lp^0}{\Lp^0}\right)^{-3/2}
\eeq we expect a large increase in the rate at which the deflection of the end of the strip increases with further increases in the strip weight $G$. 
	
\begin{figure}
	\centering
	\includegraphics[width=0.65\linewidth]{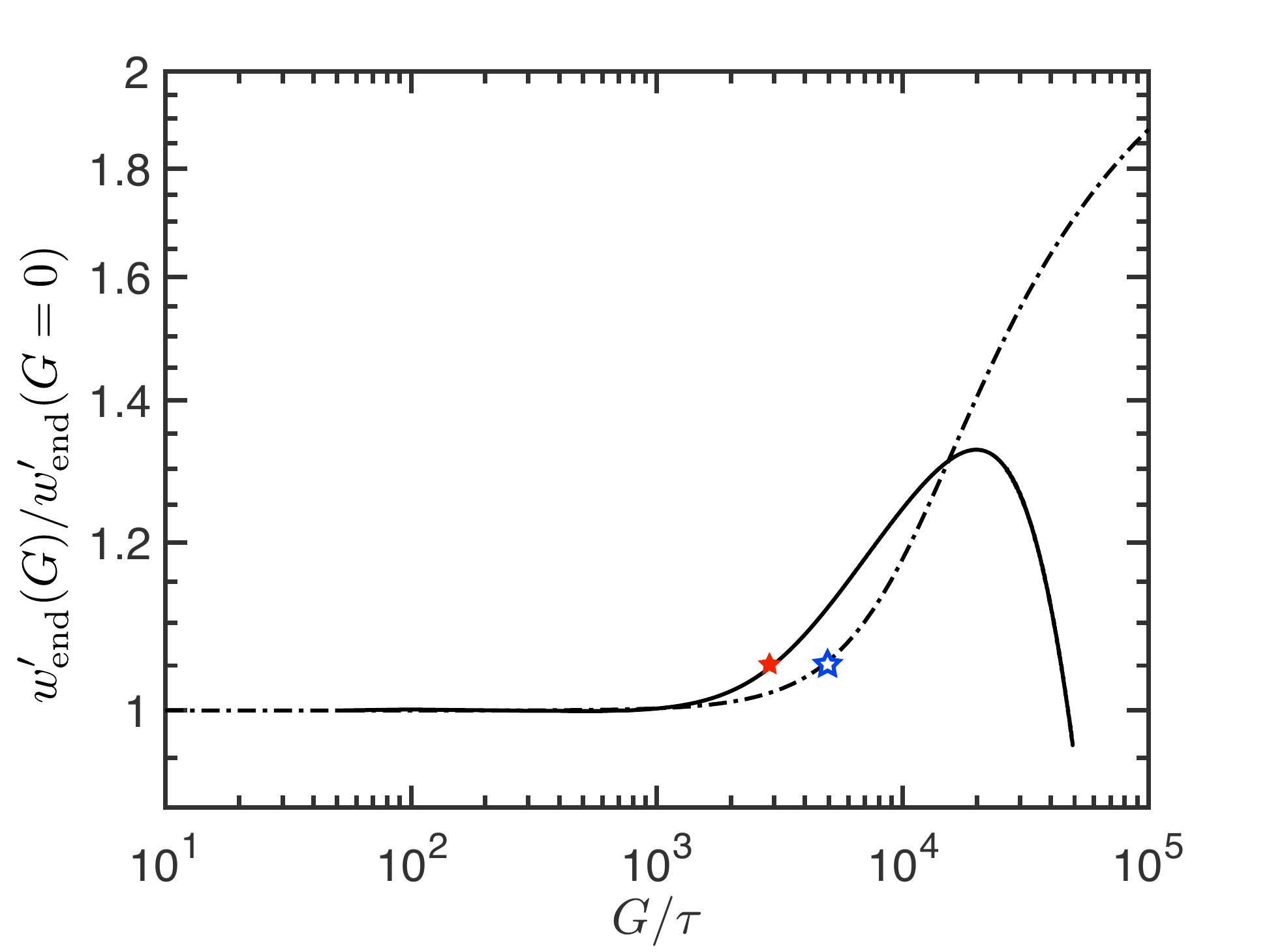}
	\caption{The buckling transition is evidenced in the  evolution of the height of the strip end, $w_{\mathrm{end}}$ as the strength of gravity changes. In particular, results from the toy model described here (dash-dotted curve) and Finite Element simulations (solid curve), described in the next section, show a marked increase beyond a critical value of $G/\tau$. In both models, buckling is defined to be where this parameter reaches a critical threshold of $1.05$, indicated by the stars. (Here $\kappa=45$, $\lambda=12$.)} \label{FIG:BucklingTransition}
\end{figure}

For this toy problem the energy, $\Utot(\tLpg)$, of eq. \eqref{eqn:scalingToy} may be readily minimized numerically  for various values of the parameters $\kappa$, $\lambda$ and $G/\tau$ and chosen values of the coefficients $\alpha_G$, $\alpha_S$ and $\alpha_B$. Here we take $\alpha_S/\alpha_B=\alphaBar^4/3$, to ensure we recover the result of Barois \emph{et al.} as $G/\tau\to0$ and $\alpha_G/\alpha_B=1/5$ for convenience. The associated end displacement $\Zend(G)$ can also be calculated, at a scaling level, leading to plots for $\Zend'(G)$ such as that shown in the dashed curve of fig.~\ref{FIG:BucklingTransition}. The results of such computations reveal a transition in $\Zend'(G)$ at a critical value of $G$ that is consistent with the scaling in \eqref{EQ:scaling}: the critical gravitational strength required to induce buckling increases as the length of the strip approaches (from above) the persistence length of the imposed curvature in the absence of gravity, $\Lp^0$. Physically, this scaling suggests that buckling is brought about by gravity reducing the length of the curved region, $\Lpg$, to the point that the uncurved region becomes long enough to bend significantly under its own weight. To test the relevance of this simple-minded approach to the buckling transition of a strip under gravity, we now turn to a series of finite element simulations.

\section{Numerical simulations}
A finite element model of the problem is implemented in the commercial software ABAQUS 6.14 (Dassault Syst\'{e}mes Simulia Corp., Johnstone, RI, USA).  The strip is represented by a mesh of 4-node, doubly-curved shell elements (S4) and using a linearly elastic Hookean material. The analysis is carried out in two steps: in the first step, the transverse curvature  is imposed at one extremity with no imposed gravity;  the numerical problem is then to determine the variation of the curvature within the strip. In the second step, the magnitude of the gravitational load is increased via a sequence of static increments. It is worth noting that numerically it is  easier  to change $G$ than to change the geometry of the strip (\emph{i.e.}~$\kappa$). No numerical stabilization is introduced.

\begin{figure}
	\centering
	\includegraphics[width=0.65\linewidth]{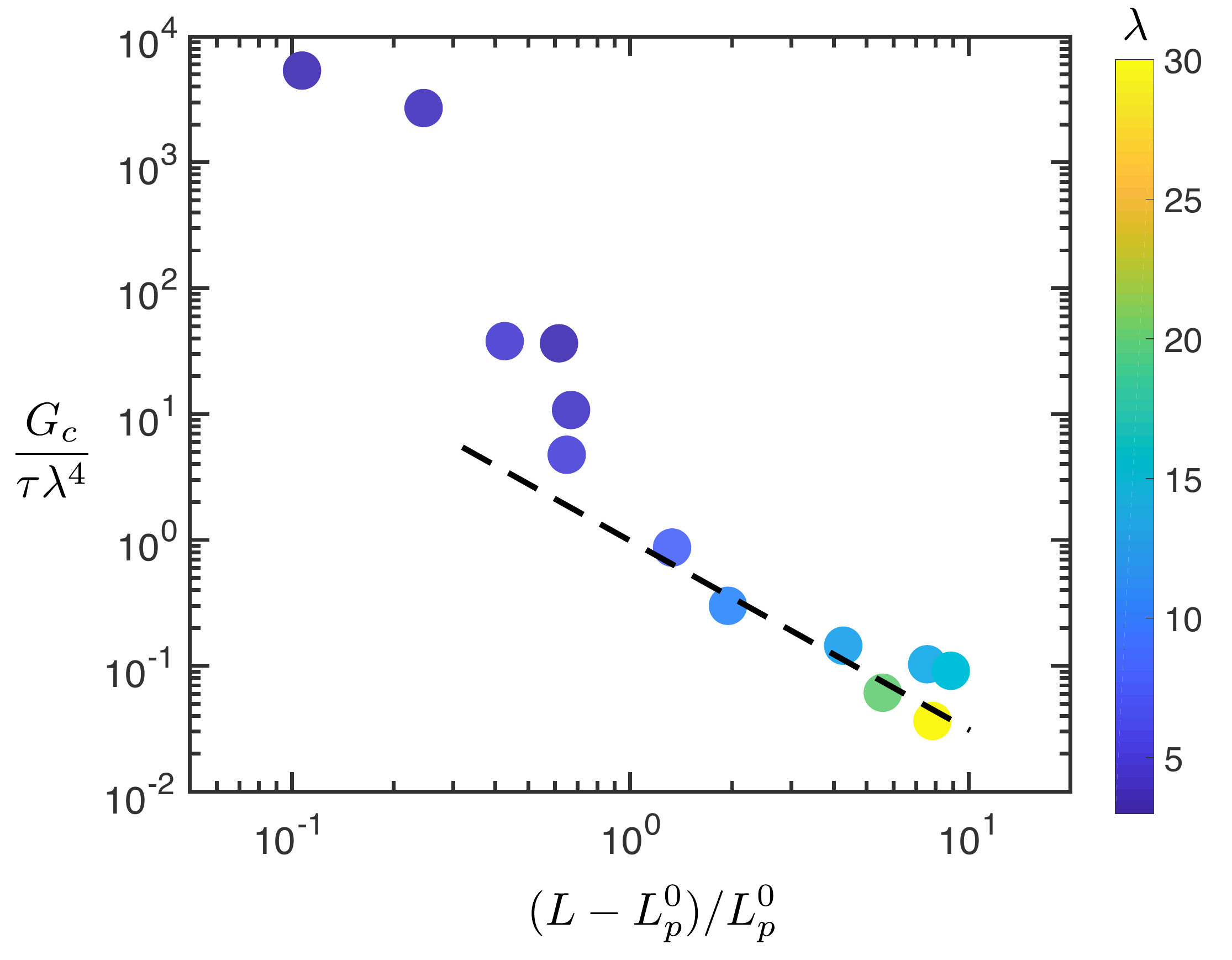}
	\caption{The critical gravitational strength, $G_c$, plotted as a function of the ratio of the uncurved: curved portions of the strip  in the absence of gravity, $(L-\Lp^0)/\Lp^0$. Numerical results are shown as points with colour coded according to the value of $\lambda$, as given in the colourbar. Plotted in this way, the prediction of the simplified model \eqref{EQ:scaling} becomes $y=x^{-3/2}$ (dashed line), and is recovered for sufficiently large values of $(L-\Lp^0)/\Lp^0$.}\label{FIG:Scaling_KappaEta}
\end{figure}

Our simulations allow us to calculate $\Zend'(G)$, with the behaviour observed qualitatively similar to that observed in the toy model (see fig.~\ref{FIG:BucklingTransition}). In these simulations, we definite the buckling event as the smallest $G$ such that $\Zend'(G)/\Zend'(G=0)=1.05$ (the stars in fig.~\ref{FIG:BucklingTransition}). We then test the scaling suggested by eq.~\eqref{EQ:scaling}: in fig.~\ref{FIG:Scaling_KappaEta}, circles indicate the numerically calculated threshold, $G_c$, while the dashed line shows the scaling of eq. \eqref{EQ:scaling}. We see that the agreement is reasonable when $L \gg L_p^0$, \emph{i.e.}~cases for which the strip is mostly uncurved  in the absence of gravity; however,  the scaling prediction \eqref{EQ:scaling} breaks down when $L-\Lp^0\lesssim\Lp^0$ --- such strips are more rigid than would be expected from \eqref{EQ:scaling}.

\begin{figure*}[h!]
	\centering
	\includegraphics[width=0.9\linewidth]{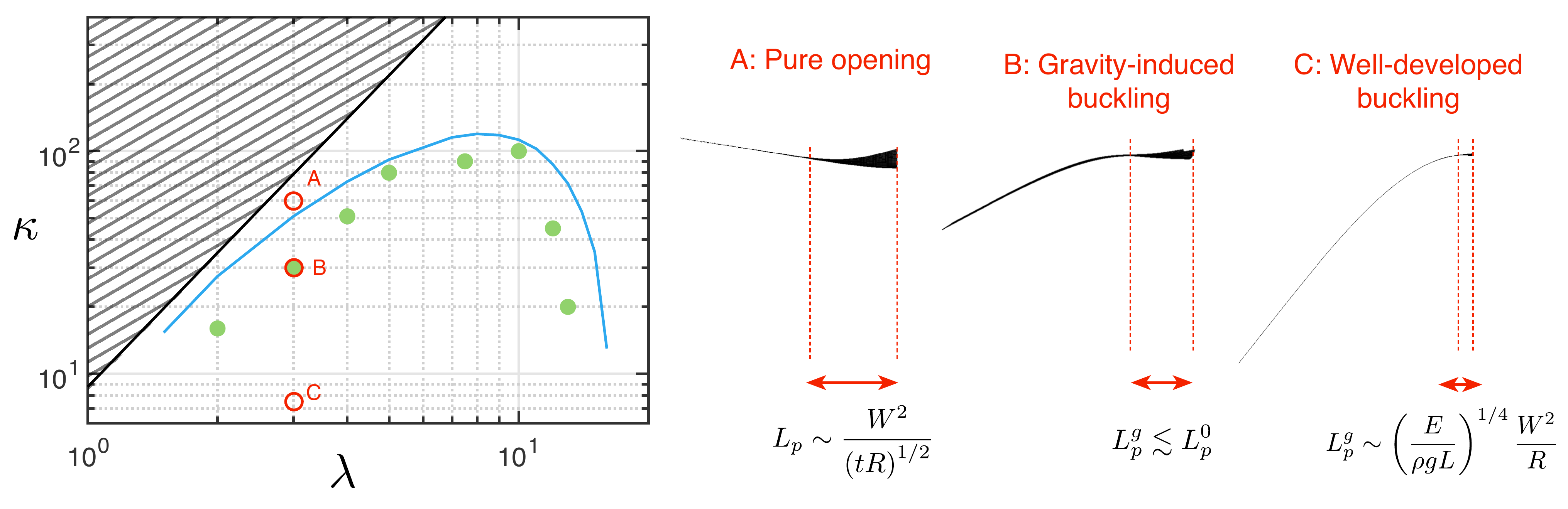}
	\caption{	Phase diagram for the imposed curvature at the onset of buckling as a function of the slenderness $\lambda$. The critical imposed curvatures $\kappac\left(\lambda, G/\tau \right)$ for  $G/\tau=2.8\pm0.1\times10^3$ obtained from FEM simulations are shown as green filled circles. The theoretical prediction from the doubly curved model is given by the continuous blue curve. To illustrate the behaviour as $\kappa$ is varied with fixed $\lambda$ and $G/\tau$, the images on the right show the strip shapes obtained from finite element simulations; shapes with $\lambda=3$, $G/\tau=2.86\times10^3$ and three values of $\kappa$ corresponding to the red open circles in the main plot.}
	\label{FIG:PhaseDiagram}
\end{figure*}

Further results for the behaviour in this critical region are shown in fig.~\ref{FIG:PhaseDiagram}. Note that in this study, we do not consider cases for which $L<\Lp^0$ (the hatched area in fig.~\ref{FIG:PhaseDiagram}).

To understand how a finite strip thickness influences the critical curvature for buckling/rigidification, fig.~\ref{FIG:PhaseDiagram} shows the qualitatively different behaviours that can be obtained for fixed values of $\lambda$, $G/\tau$ and $\nu=0$ (here, $\lambda = 3$ and $G/\tau \approx 2.86 \times10^3$). We obtain three types of behaviour as the imposed radius of curvature, $\kappa$, is varied:  (A) For strong imposed curvature (large $\kappa$),  the effect of gravity on the strip shape is negligible (variations in the transverse curvature result solely from the finite bending stiffness `uncurving' the strip); (B) Close to the buckling threshold, the strip starts to feel the effect of gravity, deforming slightly downwards, and the imposed transverse curvature decays over a shorter distance than in (A); (C) For relatively small $\kappa$, the strip exhibits large-amplitude buckling with the far end of the strip orientated parallel to the direction of gravity. In this third case the strip comprises two distinct regions: an approximately horizontal region in which the transverse curvature decays with distance along the strip and a larger, approximately vertical, region in which the strip has lost any memory of the imposed  transverse curvature, and is curved (in the longitudinal direction) solely by the effect of gravity. We  revisit the behaviour in this highly deformed region later, but focus for now on further understanding the buckling transition.

\section{Experiments}
The results of our numerical simulations suggest that the efficacy of curvature-induced rigidification is significantly enhanced for strips of comparable length to the persistence of curvature $\Lp^0$, or $\lambda$ close to $\kappa^{1/2}$. This feature that was not captured by our  analysis of the toy energy functional \eqref{eqn:scalingToy}.  To test this conclusion further, we performed a set of experiments on strips of plastic (RS Pro Shim Kit, RS Components Ltd., Northants, UK). In particular, thickness and width values in the ranges $100\leq t\leq 484$\,$\mu$m and $30\leq W\leq 91$\,mm were used, with the strip density $898\mathrm{kg\,m^{-3}}\leq\rho\leq1464\mathrm{kg\,m^{-3}}$, Young's modulus $1.0\mathrm{~GPa}\leq E\leq4.1\mathrm{~GPa}$ and Poisson's ratio $\nu=0.4$. (Young's modulus was determined via tensile tests using an Instron 3345 (Instron, Massachusetts, U.S.A)). Strips were clamped in a holder with radius of curvature $R = 4$\,cm or $R = 5$\,cm in such a way that the length of the strip $L$ could be varied ($G$ and $\lambda$ were simultaneously varied whilst maintaining a constant $\kappa$). The strip was imaged using a DSLR camera (D7000, Nikon), with a spatial resolution of $0.02$\,mm/pixel, positioned orthogonal to the major axis of the strip. The profile of the strip was extracted from the images using detection techniques developed in M{\small ATLAB} (Mathworks, Massachusetts, U.S.A). This enabled measurement of the variation in apparent thickness along the strip length and the deflection of the centreline of the strip shown in the inset to fig.~\ref{FIG:Bifurcation}. The critical length at which buckling occurred could then be determined from measurements of the displacement of the free end; the threshold was identified from abrupt changes in the displacement of the distal end that were observed as $L$ is increased. These critical conditions $\kappac(\lambda,G/\tau)$ for buckling are represented by markers in fig.~\ref{FIG:Bifurcation}, to be compared with a further theoretical analysis. 

\section{\label{SEC:BUCKLING}Doubly curved model}
\begin{figure}
	\centering
	\includegraphics[width=0.65\linewidth]{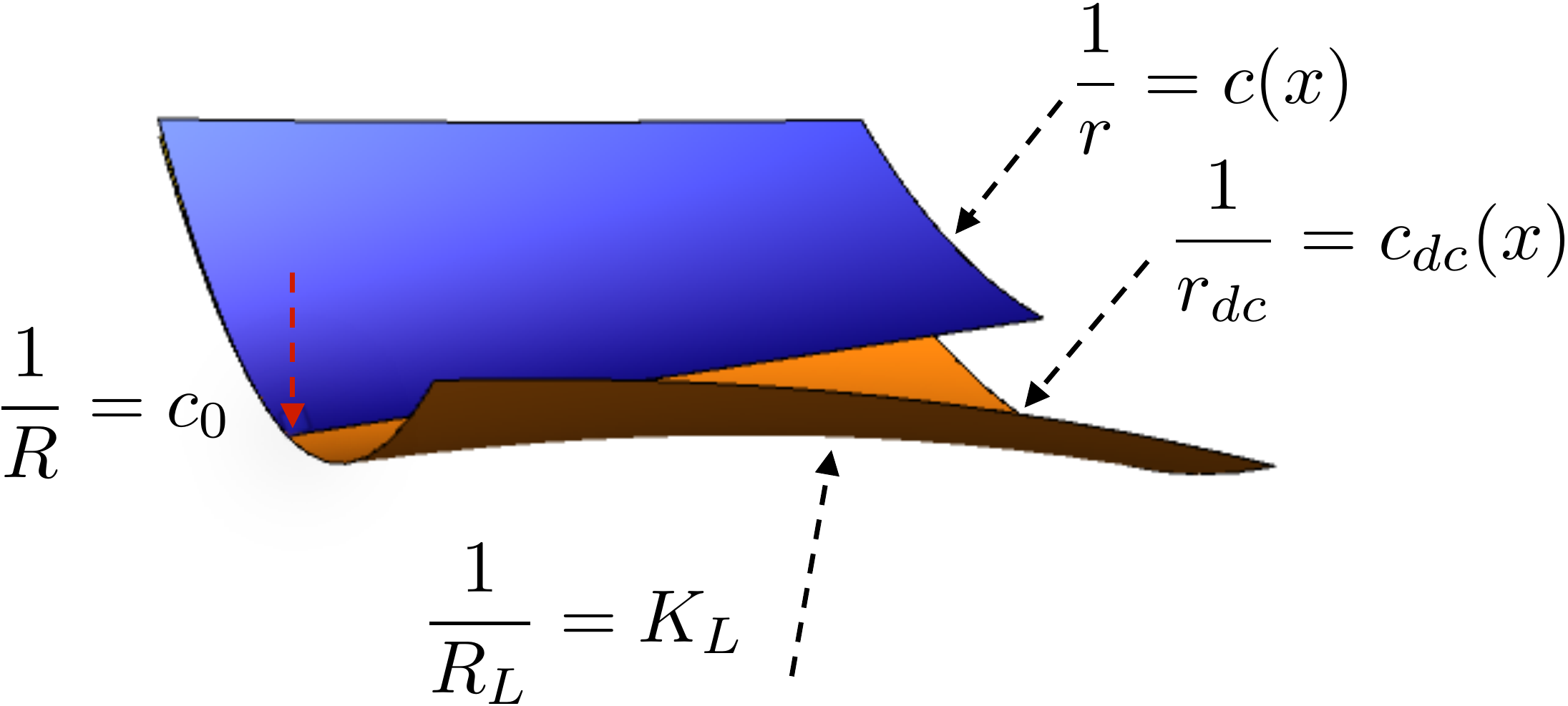}
	\caption{ The shape of a strip with a single imposed curvature (blue) is compared with the shape  of a doubly-curved strip (orange). The transverse curvature is assumed positive and the longitudinal curvature negative, \emph{i.e.} $- K_L$.} \label{FIG:DC}
\end{figure}

Returning to the buckling transition for $L-\Lp^0\lesssim\Lp^0$, we now present a more detailed analysis of an analogue system. We begin by assuming that in this limit the primary effect of   gravity is to impose a longitudinal curvature on the   transversely curved ribbon (since there is very little uncurved ribbon remaining); we therefore consider the effect of an  imposed curvature in the longitudinal (orthogonal) direction, $K_L$, (see fig.~\ref{FIG:DC}) and neglect any other effect of gravity. We shall find that there is a critical curvature $K_L^c$ at which the ribbon buckles and then estimate the buckling threshold under gravity by equating this critical curvature to that induced by the self-weight  of the strip.

For compactness,  we describe here only the main ideas while the details are presented in the Supplementary Materials \cite{SupplMat}. In the absence of gravity, we assume that the displacement field in the strip has the form $u = a_x (x)$, $v=0$ and $w = a_z(x) +\mathcal{Z}$ where $\mathcal{Z} = c(x) y^2/2$ is the out-of-plane displacement relative to the displacement of the centreline, which is in turn determined from the functions $a_z(x)$ and $a_x(x)$ \cite{Barois2014_prl}. Rescaling  curvatures by a factor $1/R$ the following dimensionless quantities can be introduced:
	\begin{equation}\label{EQ:nondimensionalization}\begin{split}
	\gamma = c R; \quad \xi = \frac{x}{L}; \quad \eta = \frac{y}{W};\quad \omega(\xi,\eta) = w(x,y) \frac{R}{W^2};\\
	\alpha_x(\xi) = a_x(x) \frac{R^2 L}{W^4}; \quad \alpha_z(\xi) = a_z(x) \frac{R}{W^2}.
	\end{split}\end{equation}
 We denote  the dimensionless transverse curvature accommodated by the ribbon after the application of the dimensionless longitudinal curvature $\kappa_L = R K_L$ by  $\gamma_{dc}(\xi) $; in general,  the transverse curvature $ \gamma_{dc}(\xi)$ is different from that of a strip opening purely under the presence of bending stiffness effects \cite{Barois2014_prl}. The relevant (dimensionless) energetic contributions are  the stretching and bending energies in this regime, which we denote by $\mathcal{U}_S^{dc}$ and $\mathcal{U}_B^{dc}$, respectively, and may be written
\beq	
\mathcal{U}_S^{dc} = 768 \kappa^2  \int_{0}^{L_p/L}\int_{-\frac{1}{2}}^{\frac{1}{2}} \left(\bar{\epsilon}^{dc} - \lambda^{-2} \bar{\epsilon}\right)^2 \upd \xi \upd \eta
\eeq and
\begin{equation}\begin{split}\label{EQ:EnergiesPS}
\mathcal{U}_B^{dc} = \int_{0}^{L_p/L}\int_{-\tfrac{1}{2}}^{\tfrac{1}{2}}\left[\left( \kappa_L + \lambda^{-2} \frac{\upd^2 \alpha_z}{\upd \xi^2} \right)^2 +(\gamma_{dc} - \gamma)^2\right] \upd \xi \upd \eta +\int_{L_p/L}^{1}\int_{-\frac{1}{2}}^{\frac{1}{2}}\kappa_L ^2 \upd \xi \upd \eta
\end{split}\end{equation}  where the stretching contribution is given by the dimensionless strain $\bar{\epsilon}^{dc} = - \eta^2 \gamma_dc \kappa_L$ minus the longitudinal strain present before the application of $\kappa_L$ and disregarding the higher order terms as in Barois \emph{et al.}\cite{Barois2014_prl,SupplMat}, as $\bar{\epsilon} = \upd \alpha_x/ \upd \xi +\left(1/2\right) (\upd \alpha_z/\upd \xi) (\upd \gamma/\upd \xi) \eta^2  +\left(1/8\right) \left(\upd \gamma / \upd \xi\right)^2 \eta^4$. Given this functional, we  compute  the transverse curvature $\gamma_{dc}(\xi)$ that minimizes the total energy $\mathcal{U}_{\mathrm{tot}}^{dc} = \left(\mathcal{U}_B^{dc} +\mathcal{U}_S^{dc}\right)$ and, in turn derive the (minimized) energy. This energy is non-convex as a function of the longitudinal curvature $\kappa_L$, so that buckling occurs \cite{Ponomarenko2012,James1981} at a critical longitudinal curvature $\kappaLc$ where  $\kappaLc( G/\tau,\lambda) = \{ \kappa_L : \partial^2 \mathcal{U}_\mathrm{tot}^{dc} / \partial \kappa_L^2 =0 \}$. 
\begin{figure}
	\centering
	\includegraphics[width=0.65\linewidth]{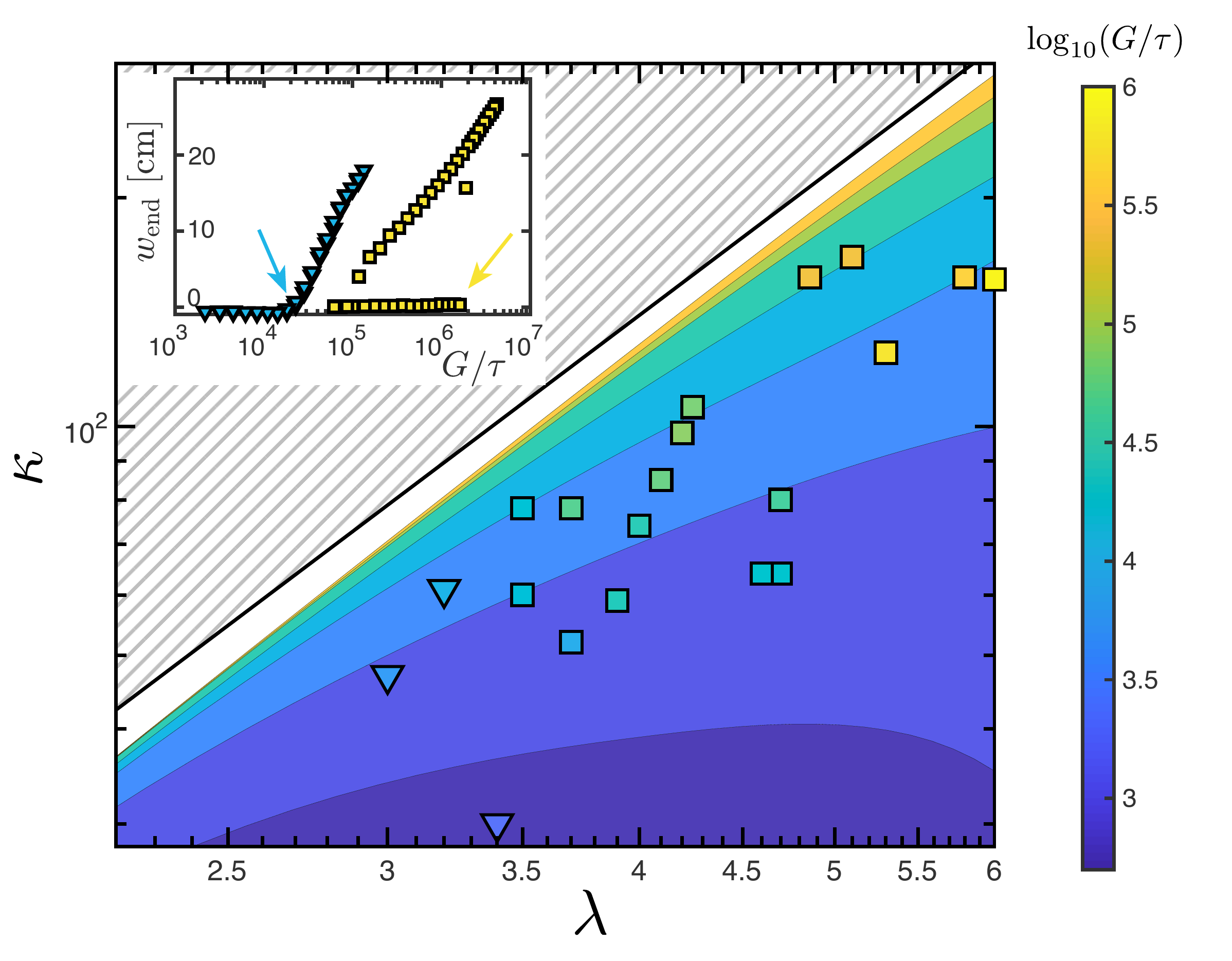}
	\caption{ 	Main figure: the critical curvature $\kappac$ at which buckling occurs for a given $G/\tau$ (with the value of $G/\tau$ represented by  the colour coding indicated in the colourbar) as a function of the strip slenderness $\lambda=L/W$. The theoretical predictions from the doubly curved model (background colour) are compared with the experimental data (markers). (Perfect agreement would lead to experimental points being camouflaged against the background colour.) Square markers represent subcritical buckling under cyclical loading, while inverted triangles represent supercritical buckling. Inset: the displacement of the free end when buckling is subcritical (yellow squares) and when buckling is supercritical (blue inverted triangles); coloured arrows indicate the point at which buckling is determined to occur for the two cases shown.}
	\label{FIG:Bifurcation}
\end{figure}

Having determined the critical imposed longitudinal curvature $\kappaLc$ for a doubly-curved strip, we estimate the averaged longitudinal curvature in a ribbon with imposed transversal curvature under the influence of gravity. Since we assumed that buckling occurs while deformations remain small, we might imagine that both the transversal curvature and the persistence length do not change substantially compared to the case without gravity, \emph{i.e.} $\Lpg \approx \Lp$ (this approach is therefore not suitable to characterize strips with $L \leq \Lp$). Within this approximation and using Mathematica (Wolfram Research Inc., Champaign, IL), it is possible to minimize the energy to derive analytical expressions for $\omega^{\mathcal{C}}_g$ and $\omega^{\mathcal{F}}_g$ so that the averaged longitudinal curvature (in the small strain approximation) in the ribbon is given by
\begin{equation}\begin{split}
\bar{\omega''} = \int_{0}^{L_p/L} \left(\frac{\upd^2 \omega^{\mathcal{C}}_g }{\upd \xi^2}\right) \upd \xi +  \int_{L_p/L}^{1} \left(\frac{\upd^2 \omega^{\mathcal{F}}_g }{\upd \xi^2}\right) \upd \xi
\left.=\frac{\upd \omega^{\mathcal{F}}_g}{\upd \xi}\right|_{\xi=1}-\left.\frac{\upd \omega^{\mathcal{C}}_g}{\upd \xi}\right|_{\xi=0}
\end{split}\end{equation}
when continuity of the first derivative at $\xi=\Lp/L$ is imposed. We take as the condition for the onset of buckling that the curvature of the strip induced by gravity, $\bar{\omega''}$, matches the critical imposed curvature at which the same strip will buckle, \emph{i.e.}
\begin{equation}\label{EQ:Threshold}
\bar{\omega''} = \kappaLc.
\end{equation}

The result of this theoretical analysis is  shown by the continuous curve in fig.~\ref{FIG:PhaseDiagram} and the background colourmap in fig.~\ref{FIG:Bifurcation}.  The utility of this theoretical description is confirmed by comparisons between its predictions and the experimental measurements shown in fig.~\ref{FIG:Bifurcation}: here the critical curvature $\kappac$ at buckling is plotted as a function of $\lambda$ for different $G/\tau$. Both experiments and theory show that, for a given slenderness,  $\lambda$, the curvature required to rigidify a strip against gravity increases with the ratio of gravitational to bending energy, $G/\tau$. A further experimental observation is that that buckling is supercritical for small $\lambda$ yet subcritical for large $\lambda$ (discussed further in the Conclusion);  even so, the predicted buckling threshold is in reasonable agreement with experiments in both cases.   We note that, this analysis, and that provided earlier for strips long compared to $\Lp^0$, give predictions for $\kappac\left(\lambda,G/\tau\right)$ when the persistence length $\Lpg$ is not significantly changed from that in the absence of gravity; we now turn to consider how the persistence length of curvature, $\Lpg$, varies far beyond the buckling threshold.

\section{\label{SEC:LpG_scaling} Behaviour well beyond threshold}
\begin{figure}[ht]
	\centering
	\includegraphics[width=0.65\linewidth]{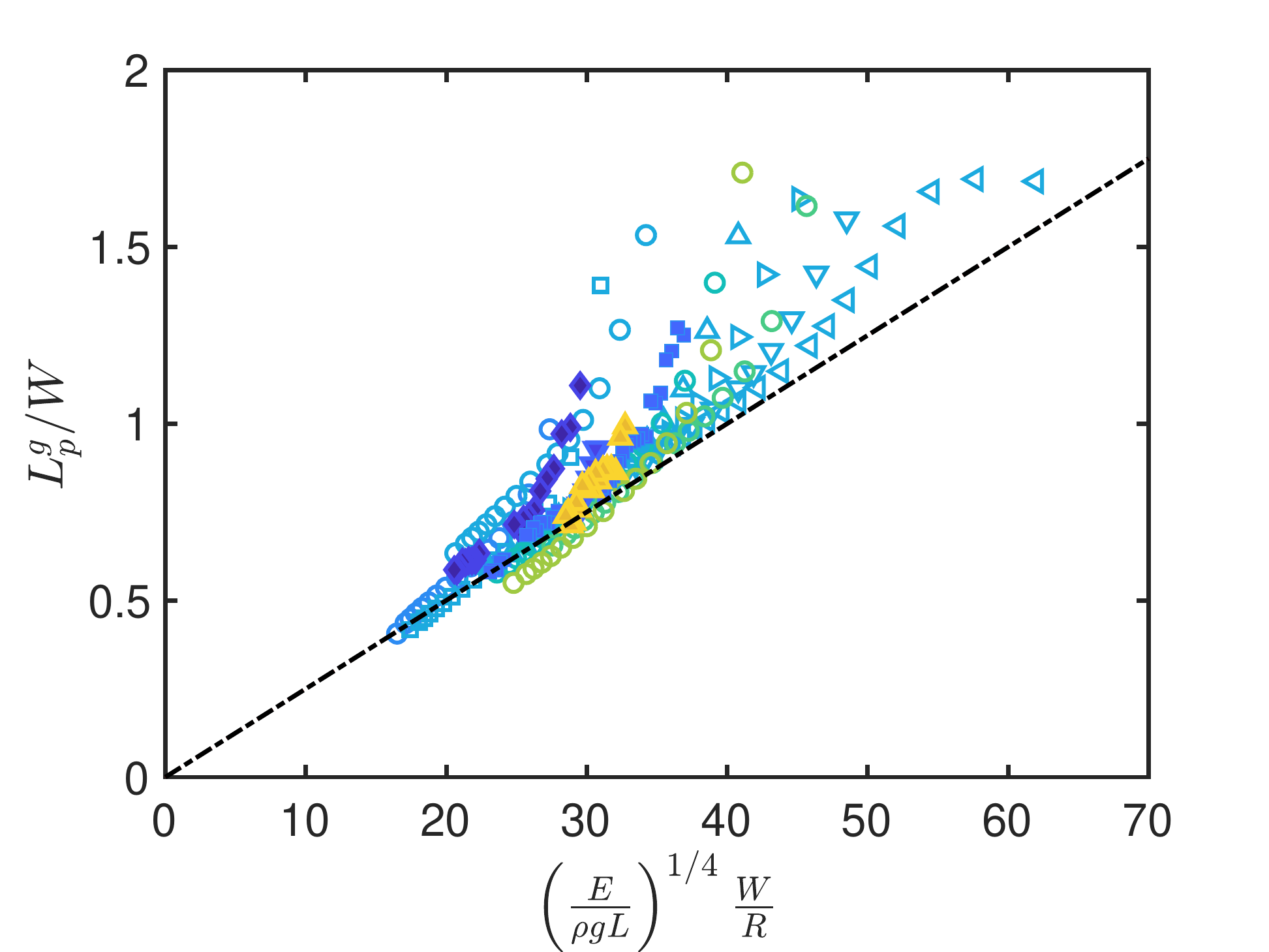}
	\caption{	The persistence length $L_p^g$ well beyond the buckling threshold, measured for different geometrical and constitutive parameters (reported in the Supplementary Material \cite{SupplMat}) in experiments (closed symbols) and finite element simulations (open symbols), collapse onto a single trend upon normalization:  the slope of the black dash-dotted line is derived as a best fit of the FEM data to be $0.025$.}\label{FIG:LpG} 
\end{figure}

Finally, we turn to the behaviour well beyond the buckling threshold; we assume again that the effect of the gravitational contribution in the domain $\mathcal{C}$ can be neglected and that $L_p^g \rightarrow 0$, so that may return to the earlier toy model. In this case, the gravitational energy of \eqref{eqn:scalingToy} must be revisited: now the displacement of the end is limited by the length of the strip, $\Zend\sim-(L-\Lpg)$, and hence $U_G\sim \rho g t W(L-\Lpg)^2$. We therefore replace the term $\alpha_G (1-\tLpg)^{5}(G/\tau)^2$ of \eqref{eqn:scalingToy} with $\alpha_G (1-\tLpg)^2(G/\tau)$. Moreover, we neglect the effect of the bending energy in eq. \eqref{eqn:scalingToy} (since $G\gg1$ and $\Lpg\ll1$, we anticipate gravity and stretching dominate instead). We see from the modified eq. \eqref{eqn:scalingToy}  that minimizing the stretching energy requires the horizontal segment of the beam to be as long as possible (increasing $\Lpg$) while the release of gravitational potential energy pushes the system to decrease $\Lpg$. Minimizing the total energy  allows us to determine the optimal persistence length of the imposed curvature:
\begin{equation}\label{EQ:normalization_LpG_long}
\frac{L_p^g}{W} \sim\lambda \tau^{1/2}G^{-1/4} \kappa
\end{equation}
where we require $L_p^g \ll L$. In terms of physical parameters, we find:
\begin{equation}\label{EQ:normalization_LpG_long_DIM}
\frac{L_p^g}{W} \sim \left(\frac{E}{\rho g L}\right)^{1/4}\frac{W}{R}.
\end{equation} This prediction is evaluated in fig.~\ref{FIG:LpG} using experiments and FEM simulations: we observe a reasonable collapse, showing that our prediction is qualitatively correct.

\section{Conclusion}
The weight of a rectangular ribbon of finite thickness, with an imposed transverse curvature at one end, acts to bend the strip against the imposed curvature. Our analysis has shown that for strips that are long, $(L-\Lp^0)/\Lp^0\gg1$, the effect of gravity is to decrease the persistence length of the imposed curvature, enhancing the bending of the flat portion of the strip. In particular, eq. \eqref{EQ:scaling} suggests in dimensional terms that the critical length of the flat portion of the strip
	\beq
	\frac{L-\Lp^0}{\leg}\sim\frac{\leg t^{1/6}}{R^{1/2}W^{2/3}}
	\eeq where $\leg=(B/\rho g t)^{1/3}$ is the length beyond which a flat cantilever bends significantly \cite{Wang1986}.
	
	As the strip length $L\searrow\Lp^0$, a more detailed study of the energy of the system is required. We note that the structure of the energy functional we derived has similarities with the one discussed by Xuan and Biggins \cite{Xuan2017} for bead-on-string formation in stretched elastic cylinders subject to superficial tension. In our case, the \emph{destabilizing} contribution is the increased applied longitudinal curvature that plays the same role that the applied stretch did in ref.~\cite{Xuan2017}; this phenomenon is mainly controlled by the competition between the two orthogonal curvatures: the increasing longitudinal bending forces the strip to open transversely, partly relieving the longitudinal stretching.  However, we have only considered strips of length $L>\Lp^0$; to our knowledge, the problem of how buckling occurs for sufficiently weak curvatures that $L<\Lp^0$ remains open.

Our experiments also suggest that the nature of the buckling changes depending on the imposed curvature. In the experiments, the buckling threshold is approached for a given $G$ by varying  $L$ cyclically, and the tip deflection exhibits hysteresis for large imposed curvature. A formal understanding of this phenomenon is beyond the scope of this paper, but we note that the transition from subcritical (denoted by squares in fig.~\ref{FIG:Bifurcation}) to supercritical (denoted by inverted triangles in fig.~\ref{FIG:Bifurcation}) occurs for $\kappa \approx 60$, and the slenderness seems to have a minor effect. The subcritical nature of the buckling presents an additional danger for pizza-eaters, since, even within the stable, rigid regime, a small perturbation may cause a catastrophic reconfiguration of the strip from an almost horizontal configuration to the buckled shape, with the distal end of the slice orientated vertically, and with very messy consequences. Aside from the  implications for eating habits, the work presented here provides a mechanical understanding of curvature-induced rigidity that might prove useful in applications such as the design of dielectric elastomer actuators with variable stiffness and tunable load capacity \cite{Li2019}. 
\vskip6pt

\section*{Acknowledgment}
This research has received funding from the European Research Council under the European Union's Horizon 2020 Programme/ERC grant agreement no. 637334. We thank M.~Gomez and O.~Kodio for discussions about this work.

\vskip2pc

\end{document}


\title{\s{Limitations of curvature--induced rigidity: How a curved strip buckles under gravity}\\
	Supplementary Material}
\author{ \textsf{Matteo Taffetani$^{1}$, Finn Box$^{1}$, Arthur Neveu$^{1}$ and Dominic Vella$^{1}$}\\ 
{\it$^{1}$Mathematical Institute, University of Oxford, UK}}

\date{\today}
\maketitle

In this Supplementary Material we provide further details of the mathematical modelling of the doubly-curved strip presented in the main text. We begin, in Section \ref{SEC:NoGravity}, by recovering the persistence length accommodated by a strip with a sufficiently large imposed curvature at one end in the absence of gravity, following Barois \emph{et al.}~\cite{Barois2014_prl}. It is instructive to \red{repeat} this analysis to introduce quantities that are used later in this document and to highlight the range of validity of the results. \red{In section \ref{SEC:DoublyCurved} we discuss the mechanics of a doubly curved strip. We begin in \S \ref{SEC:Inertia} to derive an expression for the moment of inertia of a section of a curved strip. We then} use this result, in \S\ref{sec:bend}, to present the details on the buckling of a doubly-curved strip.  In section \ref{SEC:Onset}, we \red{discuss} the procedure used to identify the onset of buckling from the finite element model of the curved strip subjected to the gravitational load, and in section \ref{SEC:Data} we present the details of the experimental settings used to \red{obtain the experimental data that is presented in} figures 6 and 8 of the main text.

\section{\label{SEC:NoGravity}Persistence length in absence of gravity in the case of sufficiently large imposed curvature}

Let us consider a rectangular strip of width $W$, length $L$ and thickness $t$. \red{The strip has Young's modulus $E$ and Poisson's ratio $\nu$.} If we introduce $u$, $v$ and $w$ as the displacements in the $x$-, $y$- and $z$-directions of a Cartesian reference system, the deformed shape achieved by the intrinsically flat strip when a transversal curvature is imposed at one of the short edge can be described using the ansatz for the displacement field proposed by Barois \emph{et al.}~\cite{Barois2014_prl} as
\begin{equation}\label{EQ:ansatz}
\begin{cases}
u(x,y) = a_x (x);\\
v(x,y) = 0;\\
w(x,y) = a_z(x)+\mathcal{Z}\left(x,y\right) =  a_z(x) + \frac{1}{2}y^2 c(x).
\end{cases}
\end{equation}
In-plane displacements are $u$ and $v$ while $w$ denotes the out-of-plane displacement with ${\cal Z}_0=W^2/(8R)$ being the apparent thickness at the clamped edge. We use the geometrically nonlinear plate theory \cite{Ventsel2001} to evaluate strains as
\begin{equation}\label{EQ:NLstrains}
\begin{cases}
\epsilon_{xx}(x,y) = \frac{\upd u}{\upd x} + \frac{1}{2}\left(\frac{\partial w}{\partial x}\right)^2 = a'_x +\frac{1}{8}y^4\left(c'\right)^2 + \frac{1}{2}y^2 c' a_z' + \left(a'_z\right)^2;\\
\epsilon_{yy}(x,y) = \frac{\upd v}{\upd y} + \frac{1}{2}\left(\frac{\partial w}{\partial y}\right)^2 = \frac{1}{2}y^2 c^2;\\
\epsilon_{xy}(x,y) = \frac{\upd u}{\upd y} + \frac{\upd y}{\upd x} +\frac{\partial w}{\partial x}\frac{\partial w}{\partial y} = \frac{1}{2}y^3 c c' + y c a'_z.
\end{cases}
\end{equation}
and curvatures as
\begin{equation}\label{EQ:NLcurvatures}
\begin{cases}
\chi_{xx}(x,y) = \frac{\partial^2 w}{\partial x^2} = a''_z + \frac{1}{2}y^2 c'';\\
\chi_{yy}(x,y) = \frac{\partial^2 w}{\partial y^2} = c;\\
\chi_{xy}(x,y) = \frac{\partial^2 w}{\partial y \partial x} = c' y.
\end{cases}
\end{equation}
where primes indicate differentiation with respect to the argument $x$. 
The dimensionless parameters that we can construct from the intrinsic length scales are
\beq
\lambda=\frac{L}{W},\quad \tau=\frac{t}{L}, \quad \kappa=\frac{W^2}{8Rt}
\eeq
where the latter is the dimensionless imposed curvature.

Following Barois \emph{et al.}~\cite{Barois2014_prl}, for large $\kappa$, the relevant energy contribution can be simplified to (where we neglect the last term when using the first equation in \eqref{EQ:NLstrains})
\begin{equation}\label{EQ:EnergiesKlarge} \begin{split}
U_{KL} = \frac{E t}{2\left(1-\nu^2\right)} \int_x \int_{-W/2}^{W/2} \epsilon_{xx}^2 \upd x \upd y + \frac{E t^3}{24\left(1-\nu^2\right)} \int_x \int_{-W/2}^{W/2} \chi_{yy}^2 \upd x \upd y =\\
\frac{1}{\left(1-\nu^2\right)}\int_x \left[\frac{E \red{t}^3 W}{2^3 \times 3}c^2 +\frac{E \red{t} W^9}{2^9 \times 3^2 \times 5^2 \times 7^2} \left(c'\right)^4\right]\upd x
\end{split}\end{equation}
where $\upd a_z/\upd x = -\left(3/56\right)W^2 \left(\upd c/\upd x\right)$ and $\upd a_x/\upd x = \left(3/4480\right) W^4 \left(\upd c/\upd x\right)^2$ have been used after minimization of the functional by varying $\upd a_z/\upd x$ and $\upd a_x/\upd x$, respectively.  Although it does not affect the final result, \red{we note that our expression for $\upd a_z/\upd x $ has}  the opposite sign to that found by Barois \emph{et al.}~\cite{Barois2014_prl}: our condition guarantees no traction in the longitudinal direction, i.e.~$\int_{-W/2}^{W/2} E t\epsilon_{xx} \upd y =0$.

We non-dimensionalize the system by letting
\begin{equation}\label{EQ:nondimensionalization}\begin{split}
\gamma = c R; \quad \xi = \frac{x}{L}; \quad \eta = \frac{y}{W};\quad \omega\red{(\xi,\eta)} = w\red{(x,y)\times} \frac{R}{W^2};\\
\alpha_x\red{(\xi)} = a_x\red{(x)} \frac{R^2 L}{W^4}; \quad \alpha_z\red{(\xi)} = a_z\red{(x)} \frac{R}{W^2}; \quad \mathcal{U}_i = U_i \frac{1}{E W h L}
\end{split}\end{equation}
with the subscript $i$ indicating the relevant energy formulation used. The energy associated with the longitudinal curvature can then be rewritten as
\begin{equation}\label{EQ:EnergiesKlargeNORM} \begin{split}
\mathcal{U}_{KL} = \frac{2^2 \tau^4 \kappa^2 \lambda^4}{\left(1-\nu^2\right)}\int_\xi \left[\frac{1}{3}\gamma\left(\xi\right)^2 +\frac{\kappa^2 \lambda^{-4}}{3^2 \times 5^2 \times 7^2} \left(\frac{\upd \gamma}{\upd \xi}\right)^4\right]\upd \xi.
\end{split}\end{equation}
Minimization of the energy, $\delta \mathcal{U}_{KL} =0$, leads to the following Euler-Lagrange equation:
\begin{equation}
-\frac{2^2 \kappa^2 \lambda^{-4}}{3 \times 5^2 \times 7^2}\left(\frac{\upd \gamma}{\upd \xi}\right)^2\frac{\upd^2 \gamma}{\upd \xi^2}+\frac{2}{3}\gamma\left(\xi\right) = 0.
\end{equation}
Multiplying this equation by $\upd \gamma/\upd \xi$ and integrating, one obtains
\begin{equation}\label{EQ:equilibrium}
\frac{5^2 \times 7^2}{2}\lambda^4 \gamma\left(\xi\right)^2 - \kappa^2 \frac{1}{2} \left(\frac{\upd \gamma}{\upd \xi}\right)^4  + \mathcal{B} = 0
\end{equation}
with $\mathcal{B}$ being the constant of integration. Here it is important to note that  we recover the solution in Barois \emph{et al.}~\cite{Barois2014_prl} only if we set $\mathcal{B} = 0$; we then find that
\begin{equation}\label{EQ:sol0}
\gamma (\xi)= \left(1 -\frac{\sqrt{35}}{2}\frac{\lambda}{\sqrt{\kappa}}\xi\right)^2
\end{equation}
after imposing $\gamma (0)=1$. Setting $\mathcal{B} = 0$ implies that this solution is valid only if the strip has, in dimensional terms, a total length $L$ larger than $L_p$. \red{(Since our analysis considers a similar energetic approach to that of ref.~\cite{Barois2014_prl}, we also consider only the case in which $L_p<L$ in this paper.)}  We can then recover the dimensionless persistence length in a form useful for the analysis in section \ref{SEC:DoublyCurved} as $L_p/L = \left(2/\sqrt{35}\right)\left(\sqrt{\kappa}/\lambda\right)$.

\section{\label{SEC:DoublyCurved}Buckling Analysis }

\red{To go beyond the simple scaling analysis presented in the main text, we need to be able to describe the effect of a changing transverse curvature on the bending of the whole strip under gravity. Since the bending stiffness $B=EI$ with $I$ the moment of inertia of the strip cross-section, it is particularly important to calculate the moment of inertia of the curved strip; we therefore consider this moment of inertia here.}

\subsection{\label{SEC:Inertia}Moment of inertia of a curved strip}

\begin{figure}[ht]
	\centering
	\includegraphics[width=5.5in]{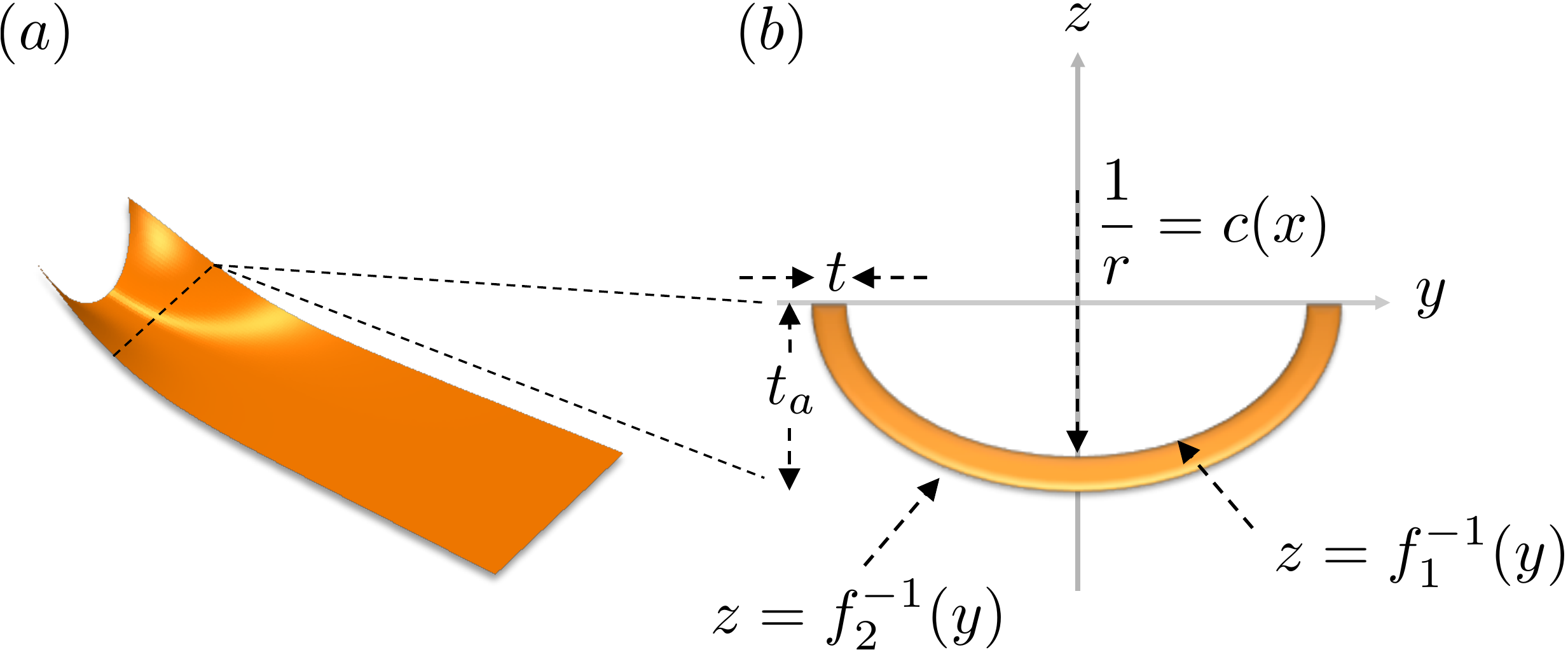}
	\caption{Sketch of the transversal cross section of the curved strip at an arbitrary longitudinal coordinate $x$.}\label{FIG:Section}
\end{figure}

Let us consider a strip of thickness $t$ and width $W$ that is naturally curved in the transverse direction with a curvature $c(x)$ that varies along the longitudinal direction. As sketched in figure ~\ref{FIG:Section}, the cross-section in the $y-z$ plane is defined as the pairs $\{\left(y,z\right) : f_1\left(y\right) = \left[\left(1/2\right) y^2 c(x) -t_a - t/2 \right]\leq z \leq (1/2) y^2 c(x) -t_a +t/2 =f_2(y), \forall y \in \left[-W/2;W/2\right]\}$, where $t_a = c(x) W^2/8$ is the \emph{apparent thickness}, i.e.~the depth in the $z$-direction, of the section at a generic longitudinal position $x$. Employing this description we assume that the thin section under analysis is given by the difference of an outer section 1 of boundary described by $z=f_1(y)$ and an inner section 2 of boundary described by $z=f_2(y)$; moreover we have $z\left(y=\pm W/2\right) =0$.

We first compute the moment of inertia $I_y$ of this section with respect to the $y$ axis. Using the additive property of the moment of inertia, we can write
\begin{equation}
I_y(x) = 2 \int_{-t_a-t/2}^{0} z^2 f_1^{-1}(z) \upd z - 2\int_{-t_a+t/2}^{0} z^2 f_2^{-1}(z) \upd z. 
\end{equation}
Expanding the results in powers of $t$ and substituting the expression for the apparent thickness $t_a$, we get
\begin{equation}
I_y(x) = \frac{1}{12} t^3 W + \frac{1}{120} t W^5 c(x)^2 + O(t^7)
\end{equation}

The areas and the static moments (with respect to the $y$ axis) of the inner and outer sections are computed as
\begin{equation}
\begin{split}
\mathcal{A}^{(1)}(x) = 2\int_{-t_a-t/2}^{0} f_1^{-1}(z) \upd z = \frac{2}{3} c(x) \left(\frac{t+2 t_a}{c(x)}\right)^{3/2}; \\
\mathcal{S}^{(1)}_y(x) = 2\int_{-t_a-t/2}^{0} z f_1^{-1}(z) \upd z = -\frac{2}{15} \left(\sqrt{2}+1\right) (t+2 t_a)^2 \left(\frac{t+2 t_a}{c(x)}\right)^{1/2};\\
\mathcal{A}^{(2)}(x) = 2\int_{-t_a+t/2}^{0} f_2^{-1}(z) \upd z = \frac{2 (2 t_a-t)^{3/2}}{3 \sqrt{c(x)}};\\ 
\mathcal{S}^{(2)}_y(x) = 2\int_{-t_a+t/2}^{0} z f_2^{-1}(z) \upd z =-\frac{2 \left(\sqrt{2}+1\right) (2 t_a-t)^{5/2}}{15 \sqrt{c(x)}}.
\end{split}
\end{equation}
We can use the additive property of the static moments and areas to compute (through subtractions) the respective quantities of our thin section, where we have already expanded the results in powers of $t$ and substituted the expression for $t_a$, as
\begin{equation}
\mathcal{A} = t W + O(t^3); \quad \mathcal{S}_y(x) = -\frac{1}{12} t W^3 c(x) + O(t^3)
\end{equation} 
Hence the $z$ coordinate of the center of mass is derived as
\begin{equation}
z_g(x) = \frac{\mathcal{S}_y(x)}{\mathcal{A}} = -\frac{1}{12} W^2 c(x) + O(t^2) .
\end{equation} 
Using the parallel axes theorem for the moment of inertia, we obtain the moment of inertia $I_0$ with respect to an axis parallel to the $y$-axis, passing through the center of mass, as
\begin{equation}\label{EQ:MomentInertia}
I_R(x) = I_y(x) - \mathcal{A} z_g(x)^2 \approx \frac{1}{12} t^3 W + \frac{1}{720} t W^5 c(x)^2.
\end{equation}

This analysis quantifies how the presence of a transverse curvature $c(x)$ stiffens a plate against the subsequent application of bending in the orthogonal direction: indeed, the moment of inertia of the transversal section increases from its base state (being $t^3 W/12$, if a rectangular section is assumed) by an amount that depends on the square of the transverse curvature, $c(x)^2$. The increase in stiffness induced by the transversal curvature, $\Delta k(x)$, measured relative to the base stiffness, $k_0(x)$, is then
\begin{equation}
\label{eqn:Pini}
\frac{\Delta k(x)}{k_0(x)} = \frac{I_R(x)-I_y(x)}{I_y(x)} = \frac{1}{60}\frac{W^4}{t^2}c(x)^2.
\end{equation}
This latter result can be directly compared with an analogous result derived by Pini \emph{et al.}~\cite{Pini2016} for a plate with an  initial longitudinal curvature. Setting the initial curvature in their equation (13) to zero, we find that \eqref{eqn:Pini} takes the same form, but allows for a transverse curvature that varies \red{in some, as yet undetermined, way} in the longitudinal direction.

\subsection{Doubly curved strip}
\begin{figure}[ht]
	\centering
	\includegraphics[width=3.5in]{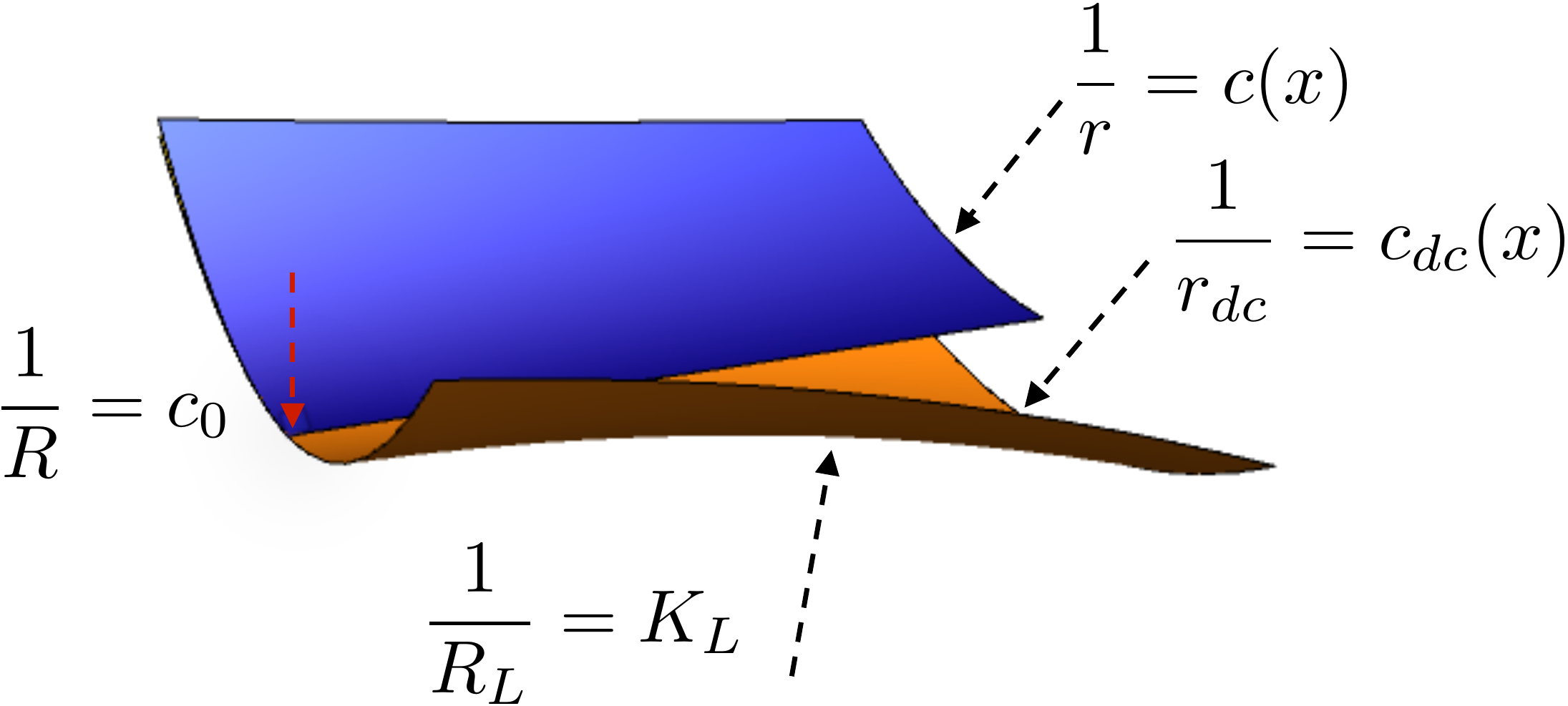}
	\caption{Doubly-curved strip (\red{orange}) compared with the shape obtained without the imposed $K_L$ (\red{blue}). We consider the transverse curvature $c_{dc}$ to be positive, while longitudinal curvature is instead negative, i.e.~$-K_L$. In the text we take into account this by placing a negative sign in front of it. }\label{FIG:PhaseGeometry}
\end{figure}

Here we consider in detail the simplified problem of a transversely curved ribbon that is forced to bend longitudinally due to the application of a constant curvature $K_L$ as shown in figure \ref{FIG:PhaseGeometry}. The approach we adopt is analogous to that presented by Ponomarenko \cite{Ponomarenko2012} in the case of a longitudinal bending of a meter tape, i.e.~a curved shell with a natural (constant in the longitudinal direction) transversal curvature. \red{However, we allow a spatially varying transverse curvature, in contrast to the constant transverse curvature of Ponomarenko \cite{Ponomarenko2012} .}

If $c_{dc}(x) \neq c(x)$ denotes the dimensional transverse curvature accommodated by the ribbon after the application of $K_L$ (while $c(x) = \left(1/R\right)\left(1-x/L_p\right)^2$ is the dimensional transverse curvature prior to the application of $K_L$), the out-of-plane displacement in this doubly-curved system may be rewritten as 
\begin{align}\label{EQ:wdc}
\begin{cases}
w_{dc}(x,y) = -\frac{1}{2}y^2 c_{dc}(x) - \frac{1}{2}K_L x^2  &\mathrm{if} \quad x \leq L_p\\
w_{dc}(x) = - \frac{1}{2}K_L x^2  &\mathrm{if} \quad x > L_p
\end{cases}
\end{align}

If we consider the longitudinal and transversal changes in curvature in the two domains, we can construct the bending contribution to the energy functional as 
\begin{equation}\begin{split}\label{EQ:EnergiesPS_bending}
U_B^{dc} = \frac{E t^3}{24\left(1-\nu^2\right)} \int_{-L_p}^{0}\int_{-W/2}^{W/2}\left[\left( - K_L - \frac{\upd^2 a_z}{\upd x^2} \right)^2 + \left(c_{dc} - c\right)^2\right] \upd x \upd y\\
+ \frac{E t^3}{24\left(1-\nu^2\right)} \int_{0}^{L-L_p} \int_{-W/2}^{W/2} K_L ^2 \upd x\,\upd y.
\end{split}\end{equation}
 The longitudinal curvature induces a differential strain between the centreline and the long edge that can be quantified as $\epsilon^{dc}(x) = -y^2 c_{dc}(x) K_L$. Superimposing this strain onto the longitudinal strain expressed in the first line of equation \eqref{EQ:NLstrains} (again neglecting the last square term) we define the relevant stretching contribution for the energy as
\begin{equation}\begin{split}\label{EQ:EnergiesPS_stretching}
U_S^{dc} = \frac{E t}{2\left(1-\nu^2\right)} \int_{-L_p}^{0} \int_{-W/2}^{W/2} \left(\epsilon^{dc} - \epsilon\right)^2 \upd x\, \upd y.
\end{split}\end{equation}
The stationarity of the total energy $U_{tot}^{dc} = \left(U_B^{dc} +U_S^{dc}\right)$ with respect to $c_{dc}(x)$ is guaranteed when the transversal curvature changes as
\begin{equation}\label{EQ:TransverseCurvatureGdim}
c_{dc}(x)= \frac{20 t^2 c(x)}{20 t^2+3 K_L^2 W^4}
\end{equation}
As intuitively expected, the longitudinal curvature tends to open (`uncurve') the strip in the transverse direction, i.e $c_{dc}(\bar{x})<c(\bar{x}), \forall \bar{x}<L_p$, thus reducing the longitudinal strain in the ribbon, $\epsilon^{dc}$. In the present form, this energy functional allows the transverse curvature $c_{dc}(x)$ to vary everywhere in the domain, while our physical problem requires $c_{dc}(0)=1/R$ to be fixed. To obtain a more precise description, the term related to the change in curvature in the longitudinal direction in equation \eqref{EQ:EnergiesPS_bending} should be replaced with $\left( - K_L -\upd^2 c_{dc}/\upd x^2 - \upd^2 a_z/\upd x^2 \right)^2$ and the corresponding differential equation should be derived: nonetheless, the error introduced by this simplification is limited to a layer close to $x=0$ and does not affect the results presented in the main text. 

For completeness, we rewrite the energy in dimensionless form:
\begin{equation}\begin{split}\label{EQ:EnergiesPS}
\mathcal{U}_{tot}^{dc} = \int_{-\frac{L_p}{L}}^{0}\int_{-\frac{1}{2}}^{\frac{1}{2}}\left[\left( - \kappa_L - \lambda^{-2} \frac{\upd^2 \alpha_z}{\upd \xi^2} \right)^2 + \left(\gamma_{dc} - \gamma\right)^2\right] \upd \xi\, \upd \eta +\int_{0}^{1-\frac{L_p}{L}}\int_{-\frac{1}{2}}^{\frac{1}{2}}\kappa_L ^2 \upd \xi\, \upd \eta \\
+ 768 \kappa^2  \int_{-\frac{L_p}{L}}^{0}\int_{-\frac{1}{2}}^{\frac{1}{2}} \left(\bar{\epsilon}^{dc} - \lambda^{-2} \bar{\epsilon}\right)^2 \upd \xi\, \upd \eta
\end{split}\end{equation}
where the stretching contribution is given by the dimensionless strain $\bar{\epsilon}^{dc} = - \eta^2 \gamma_{dc} \kappa_L$ and  $\bar{\epsilon} = \upd \alpha_x/ \upd \xi +\left(1/2\right) \upd \alpha_z/\upd \xi \upd \gamma/\upd \xi \eta^2  +\left(1/8\right) \left(\upd \gamma / \upd \xi\right)^2 \eta^4$, with $\upd \alpha_z/\upd \xi= -\left(3/56\right) \upd \gamma/\upd\xi$ and  $\upd \alpha_x/\upd \xi= -\left(3/4480\right) \left(\upd \gamma/\upd\xi\right)^2$. Thus, the dimensionless transversal curvature that minimizes this functional is
\begin{equation}\label{EQ:TransverseCurvatureG}
\gamma_{dc}(\xi)= \frac{5 \gamma(\xi)}{48 \kappa^2 \kappa_L^2+5}
\end{equation}
and the relative energy is 
\begin{equation}\begin{split}
\mathcal{U}^{dc}_{tot} = \int_0^{L_p/L} \left[\left(\gamma-\frac{5 \gamma}{48 \kappa^2 \kappa_L^2+5}\right)^2+\left(\kappa_L-\frac{3}{56} \frac{\upd^2 \gamma}{\upd \xi^2} \lambda^{-2}\right)^2\right. \\
\left.+ \frac{240 \gamma^2 \kappa^2 \kappa_L^2}{\left(48 \kappa^2 \kappa_L^2+5\right)^2}+\frac{1}{3675}\left(\frac{\upd \gamma}{\upd \xi}\right)^4 \kappa^2 \lambda^{-4} \right]\upd \xi + \kappa_L^2 \left(1-\frac{L_p}{L}\right)
\end{split}\end{equation}

Inspection of this functional reveals that the non-convex nature of this energy results from the decrease in the transverse curvature as $\sim \left[5+48 (\kappa_L\kappa)^2\right]^{-1}$, in contrast with the increase of longitudinal bending energy that occurs when we increase the applied $\kappa_L$.

The second derivative of the energy reads
\begin{equation}
\frac{\partial^2 \mathcal{\red{U}}_{tot}^{dc}}{\partial \kappa_L^2} = \frac{192 \kappa^{5/2} \left(5-144 \kappa^2 \kappa_L^2\right)}{\sqrt{35} \lambda \left(48 \kappa^2 \kappa_L^2+5\right)^3}+2
\end{equation}
where we used the definition of $\gamma$ in equation \eqref{EQ:TransverseCurvatureG} and of $L_p/L$ (as in section 1) to compute the integrals. The critical longitudinal curvature $\left(\kappa_L\right)_{c}(\kappa, \lambda)$ at which the transversely curved ribbon buckles is the smallest positive solution of
\begin{equation}
\frac{\partial^2 \mathcal{\red{U}}_{tot}^{dc}}{\partial \kappa_L^2} = 0,
\end{equation} \red{since this condition corresponds to where the moment required to impose this curvature, $M\propto\partial \mathcal{U}_{tot}^{dc}/\partial \kappa_L$, changes sign.}

\subsection{\label{sec:bend}Bending of a beam with non-homogeneous moment of inertia}

Let us recall that for a strip with Young's modulus $E$ and density $\rho$, we can extract the elastogravitational length $\leg$ by balancing the bending contribution through the bending modulus of a beam of rectangular cross section $B = E t^3 W/12$ and the gravitational contribution as $\rho g t W$,  where $g$ denotes the gravitational acceleration, so that $\leg=(B/\rho g t)^{1/3}$. Defining the gravity parameter
\beq
G= \frac{12 \rho g L^3}{E t^2}.
\eeq
we  consider here the dimensionless functional that describes the bending induced by gravity of an elastic beam with  inhomogeneous moment of inertia. Since this functional is derived assuming that the strip is wide, we retain the factor $\left(1-\nu^2\right)$ for the bending energy \cite{AudolyPomeau} and we write it as
\begin{equation}\label{EQ:gravity}
\begin{split}
\mathcal{U}_{beam}  = \frac{\kappa^2 \tau^4 }{3\left(1-\nu^2\right)} \int_\xi 
{\cal I}_R \left(\frac{\upd^2 \omega_{g}}{\upd \xi^2}\right)^2 \upd \xi +
\frac{2 G \kappa \tau^3}{3} \int_{\xi} \omega_{g} \upd \xi
\end{split}
\end{equation}
where ${\cal I}_R= 1+ (16/15) \kappa^2 \gamma(\xi)^2$ is the dimensionless moment of inertia (see equation \eqref{EQ:MomentInertia} in dimensional form), while $\omega_{g}(\xi)$ identifies the out-of-plane displacement of the centre of mass.

As in the problem of a curved ribbon in the absence of gravity, we anticipate  two domains: a transversely curved region of length $L_p^g \neq L_p(G=0)$ denoted as $\mathcal{C} = \left\{x: 0< x < L_p^g\right\}$, and a transversely flat region $\mathcal{F} = \left\{x: L_p^g<x<L\right\}$, in which the ribbon may solely be curved in the longitudinal direction. If we assume that the persistence length does not change substantially with respect to the case without gravity (subscript/superscript $g$ refer to quantities under the effect of the gravitational field), i.e.~$\gamma_g \approx \gamma$ and $L_p^g/L \approx L_p/L$, minimization of the functional in equation \eqref{EQ:gravity} in the two domains is obtained by solving the following dimensionless system:

\begin{equation}\begin{split}\label{EQ:EquilibriumEB}\begin{cases}
\frac{\kappa \tau}{G \left(1-\nu^2\right)}\frac{\upd^2}{\upd \xi^2} \left({\cal I}_R \frac{\upd^2}{\upd \xi^2} \omega^{\mathcal{C}}_g(\xi) \right) + 1 = 0 \quad\quad 0<\xi<\frac{L_p}{L}\\
\frac{\kappa \tau}{G \left(1-\nu^2\right)} \frac{\upd^4}{\upd \xi^4} \omega^{\mathcal{F}}_g(\xi)+ 1 = 0 \quad\quad \frac{L_p}{L}<\xi<1
\end{cases}\end{split}\end{equation}
To discriminate between the two domains, we replace $\omega_{g}$ with $\omega^{\mathcal{C}}_g$ in the first equation \red{(since this region is transversely curved)} and $\omega^{\mathcal{F}}_g$ in the second one \red{(since this region is transversely flat)}, recalling that ${\cal I}_R=1$ \red{in the flat region}. The set of natural boundary and essential matching conditions required to close the system are:
\begin{equation}\begin{split}\label{EQ:ConditionsEB}\begin{cases}
\omega^{\mathcal{C}}_g(0) = \frac{\upd }{\upd \xi} \omega^{\mathcal{C}}_g |_{\xi=0} = 0\\ 
\frac{\upd^2 }{\upd \xi^2} \omega^{\mathcal{F}}_g |_{\xi=1} = \frac{\upd^3 }{\upd \xi^3} \omega^{\mathcal{F}}_g |_{\xi=1} = 0\\
\frac{\upd^j }{\upd \xi^j} \omega^{\mathcal{C}}_g |_{\xi=L_p/L} = \frac{\upd^j }{\upd \xi^j}\omega^{\mathcal{F}}_g |_{\xi=L_p/L} \quad j \in [0,1,2,3]
\end{cases}\end{split}\end{equation}
The averaged longitudinal curvature (small strain approximation) in the ribbon is
\begin{equation}\begin{split}
\bar{\omega''} = \int_{0}^{L_p/L} \left(\frac{\upd^2 \omega^{\mathcal{C}}_g }{\upd \xi^2}\right) \upd \xi +  \int_{L_p/L}^{1} \left(\frac{\upd^2 \omega^{\mathcal{F}}_g }{\upd \xi^2}\right) \upd \xi= \left.\frac{\upd \omega^{\mathcal{F}}_g}{\upd \xi}\right|_{\xi=1}-\left.\frac{\upd \omega^{\mathcal{C}}_g}{\upd \xi}\right|_{\xi=0}.
\end{split}\end{equation}
when continuity of the first derivative at $\xi=L_p/L$ is imposed. The condition for the stability threshold is given by the solution of
\begin{equation}\label{EQ:Threshold}
\bar{\omega''} = \left(\kappa_L\right)_c
\end{equation}
which can be solved numerically.

\section{\label{SEC:Onset}Detection of the Buckling point}

\begin{figure}[h!]
	\centering
	\includegraphics[width=0.95\linewidth]{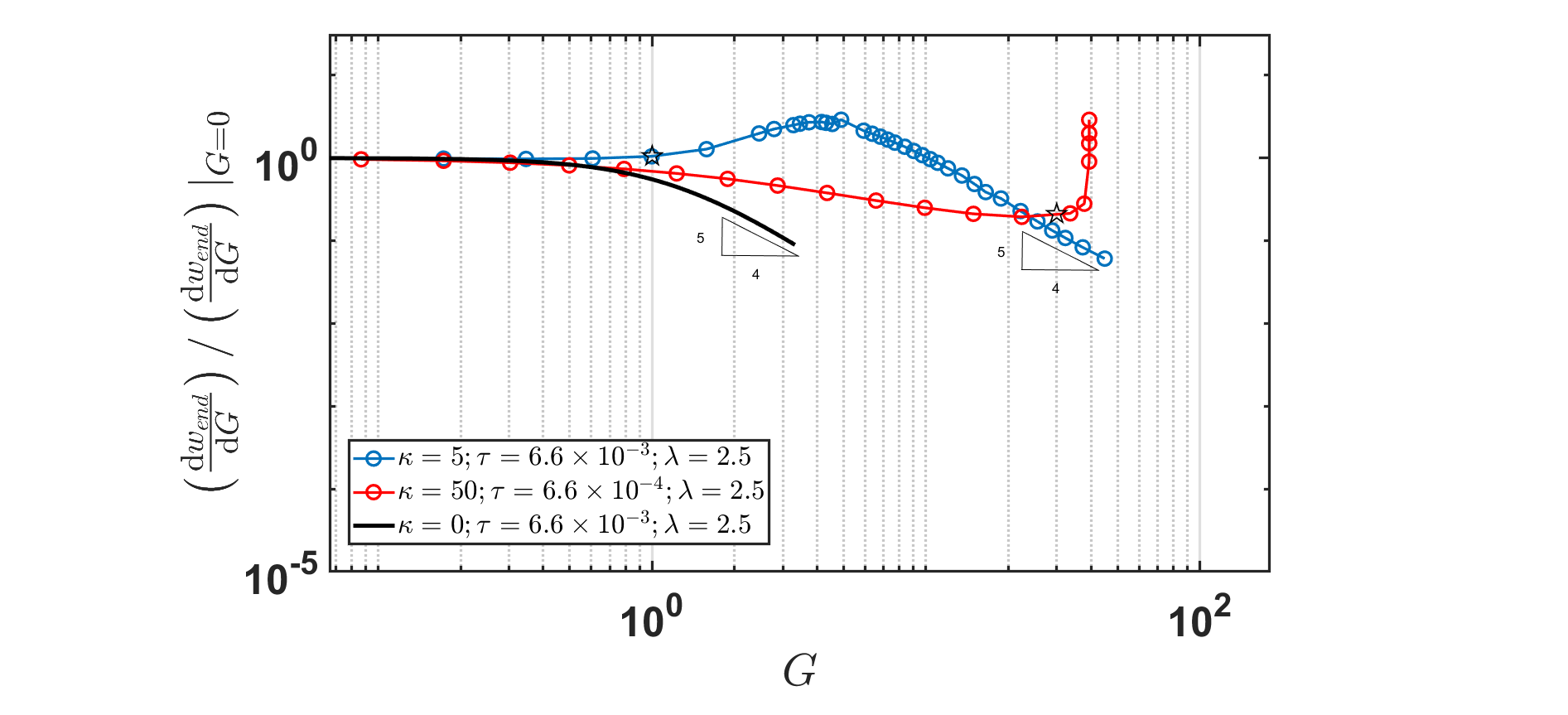}
	\caption{Derivative of the out-of-plane displacement, $\upd w_{end}/\upd G$, normalized by its value at $G=0$ as a function of $G$ for three representative cases. A cantilever with zero transverse curvature bends continuously with increasing $G$ such that $\upd w_{end}/\upd G$ monotonically decreases to $\sim G^{-5/4}$ for large $G$. Buckling is instead denoted by a region where $\upd w_{end}/\upd G$ increases: in the cases shown here, buckling occurs at $G \approx 1$ for $\kappa=5$ and $G\approx 30$ for $\kappa = 50$ \red{(stars identify these points)}.}\label{FIG:Onset}
\end{figure}

To identify the onset of buckling in the finite element simulations, we examine the derivative of the out-of-plane displacement at the short free end, $w_{end}=\omega(\xi=1)$, with respect to $G$.
We recognize the presence of two limiting regimes: (i) \red{for weak gravity,} $G\red{\ll1}$, the end deflection scales as in the bending of a Euler beam $ w_{end} \sim G$; (ii) \red{for strong gravity} $G \red{\gg1}$, the beam deformation is large\red{. In this regime} the transversely flat domain $\mathcal{F}$ is orientated parallel to the direction of gravity and $\upd w_{end} / \upd G \sim G^{-5/4}$ since we expect $w_{end} \approx L -L_p^g$ and $L_p^g \sim W \lambda \tau^{1/2}G^{-1/4} \kappa$, as derived in the main text. These two regimes are smoothly connected if the deformation path is characterized by a continuous bending, where the slope of $\upd w_{end} / \upd G$ monotonically decreases for increasing $G$. The presence of a non-monotonic behaviour is, instead, indicative of buckling. In figure \ref{FIG:Onset} we present two examples from finite element simulations to show how the quantity  $\upd w_{end} / \upd G$ qualitatively behaves for different sets of the geometrical parameters. As a reference, we also consider the bending of a transversely flat strip (i.e.~a cantilever beam) in black. In this analysis the Poisson's ratio is set as $\nu=0$. \red{To identify buckling, we use the criterion $\left(\upd w_{end} / \upd G\right)/\left(\upd w_{end} / \upd G\right)_{G=0} = 1.05$ when this parameter only increases monotonically from $G=0$ (see blue curve in fig.~\ref{FIG:Onset}); in other cases (typically when $L\sim L_p^0$) this function has a clear turning point (as visible in  the red curve of fig.~\ref{FIG:Onset}) and we identify buckling at the local minimum. }
\vskip2pc
\clearpage
\section{\label{SEC:Data}Experimental Parameters}
\begin{figure}[ht]
	\centering
	\includegraphics[width=0.65\linewidth]{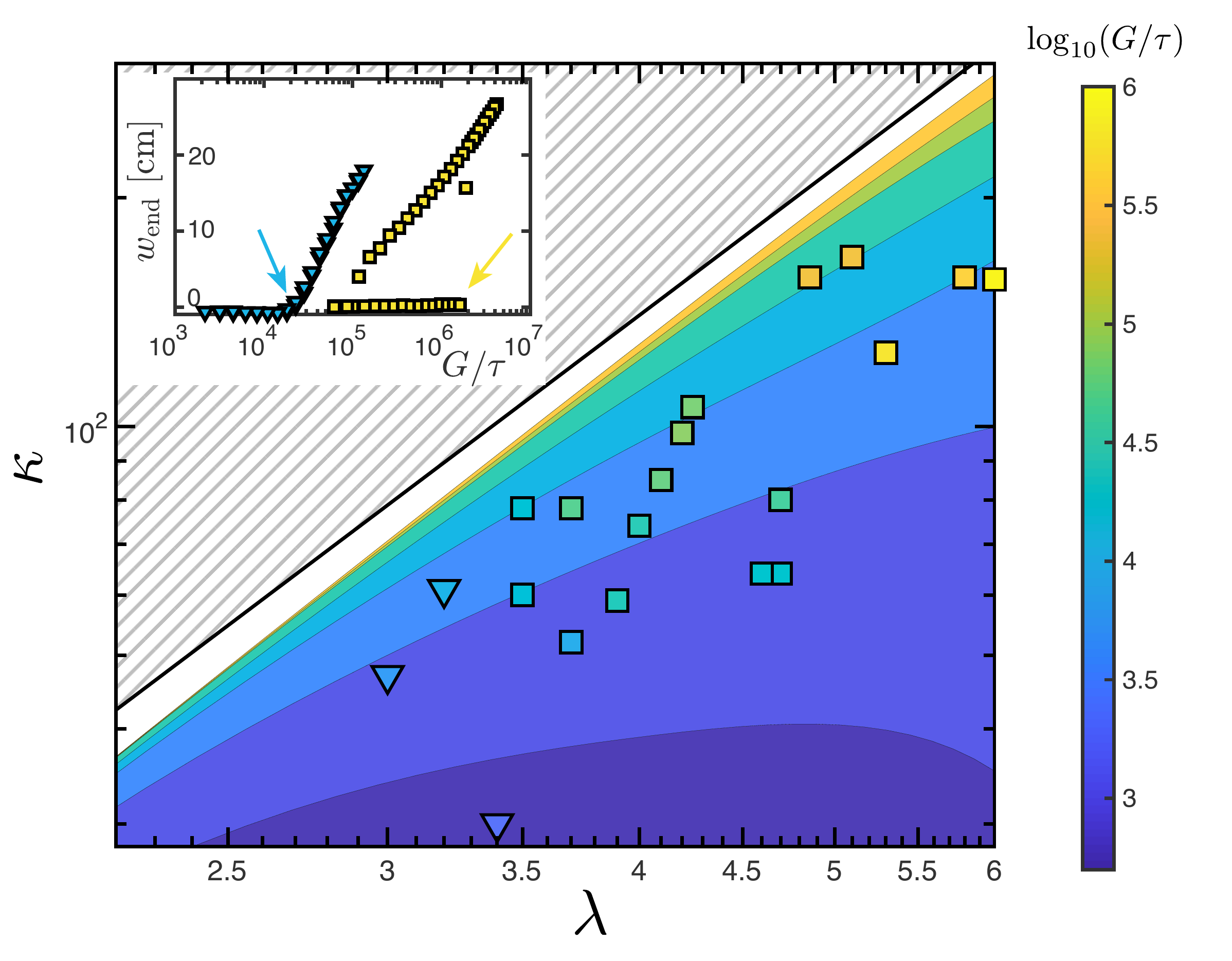}
	\caption{Main figure: the critical curvature $\kappa_c$ at which buckling occurs for a given $G/\tau$ (with the value of $G/\tau$ represented by  the colour coding indicated in the colourbar) as a function of the strip slenderness $\lambda=L/W$. The theoretical predictions from the doubly curved model (background colour) are compared with the experimental data (markers). (Perfect agreement would lead to experimental points being camouflaged against the background colour.) Square markers represent subcritical buckling under cyclical loading, while inverted triangles represent supercritical buckling. Inset: the displacement of the free end when buckling is subcritical (yellow squares) and when buckling is supercritical (blue inverted triangles); coloured arrows indicate the point at which buckling is determined to occur for the two cases shown.}
	\label{FIG:Bifurcation}
\end{figure}

In this section we record the parameter values used in the experiments and finite element simulations. In table \ref{TAB:Figure5} we list the experimental parameter used to plot figure 6 in the main paper, reported here as figure \ref{FIG:Bifurcation}, together with the marker shapes and colours used in this plot.

\definecolor{marker1}{RGB}{248,223,26}
\definecolor{marker2}{RGB}{246,228,38} %
\definecolor{marker3}{RGB}{253,205,51} %
\definecolor{marker4}{RGB}{241,187,54} %
\definecolor{marker5}{RGB}{241,187,54} %
\definecolor{marker6}{RGB}{131,205,90} %
\definecolor{marker7}{RGB}{110,205,102} %
\definecolor{marker8}{RGB}{92,205,118} %
\definecolor{marker9}{RGB}{77,205,131} %
\definecolor{marker10}{RGB}{61,202,143} %
\definecolor{marker11}{RGB}{41,195,167} %
\definecolor{marker12}{RGB}{28,195,174} %
\definecolor{marker13}{RGB}{0,187,197} %
\definecolor{marker14}{RGB}{0,187,197} %
\definecolor{marker15}{RGB}{3,184,205} %
\definecolor{marker16}{RGB}{8,182,210} %
\definecolor{marker17}{RGB}{31,166,228} %
\definecolor{marker18}{RGB}{38,159,233} %
\definecolor{marker19}{RGB}{45,135,240} %
\definecolor{marker20}{RGB}{72,90,248} %

\begin{table}[H]
	\begin{center}
		\begin{tabular}{cccccccc}
			$W$ [cm]& $R$ [cm] & $t$ [$\mu$\,m] & $E$ [G\,Pa] & $\rho$ [kgm$^{-3}$] & $\kappa$ & $G/\tau$ & Marker  \\
			$	5.0	$ & $	4.0	$ & $	50	$ & $	3.9	$ & $	1464	$ & $	156	$ & $	2.9e6	 $ & \textcolor{marker1}{$\blacksquare$}	\\
			$	5.0	$ & $	5.0	$ & $	50	$ & $	3.9	$ & $	1464	$ & $	125	$ & $	1.7e6	 $ & \textcolor{marker2}{$\blacksquare$}	\\
			$	7.3	$ & $	4.0	$ & $	106	$ & $	4.1	$ & $	1331	$ & $	157	$ & $	1.0e6	 $ & \textcolor{marker3}{$\blacksquare$}	\\
			$	7.3	$ & $	4.0	$ & $	106	$ & $	4.1	$ & $	1464	$ & $	157	$ & $	5.6e5	 $ & \textcolor{marker4}{$\blacksquare$}	\\
			$	7.3	$ & $	4.0	$ & $	100	$ & $	5.7	$ & $	1380	$ & $	167	$ & $	5.5e5	 $ & \textcolor{marker5}{$\blacksquare$}	\\
			$	5.0	$ & $	4.0	$ & $	80	$ & $	4	$ & $	1415	$ & $	98	$ & $	1.6e5	 $ & \textcolor{marker6}{$\blacksquare$}	\\
			$	6.0	$ & $	4.0	$ & $	106	$ & $	4.1	$ & $	1331	$ & $	106	$ & $	1.4e5	 $ & \textcolor{marker7}{$\blacksquare$}	\\
			$	6.0	$ & $	5.0	$ & $	106	$ & $	4.1	$ & $	1331	$ & $	85	$ & $	1.2e5	 $ & \textcolor{marker8}{$\blacksquare$}	\\
			$	5.0	$ & $	5.0	$ & $	80	$ & $	4	$ & $	1415	$ & $	78	$ & $	9.5e4 $ & \textcolor{marker9}{$\blacksquare$}	\\
			$	9.1	$ & $	5.0	$ & $	260	$ & $	4	$ & $	1400	$ & $	80	$ & $	7.8e4	 $ & \textcolor{marker10}{$\blacksquare$}	\\
			$	5.0	$ & $	4.0	$ & $	106	$ & $	4.1	$ & $	1331	$ & $	74	$ & $	5.1e4 $ & \textcolor{marker11}{$\blacksquare$}	\\
			$	5.0	$ & $	5.0	$ & $	106	$ & $	4.1	$ & $	1331	$ & $	59	$ & $	4.6e4	 $ & \textcolor{marker12}{$\blacksquare$}	\\
			$	7.3	$ & $	4.0	$ & $	260	$ & $	4	$ & $	1400	$ & $	64	$ & $	3.2e4	 $ & \textcolor{marker13}{$\blacksquare$}	\\
			$	7.3	$ & $	4.0	$ & $	260	$ & $	4	$ & $	1400	$ & $	64	$ & $	3.0e4	 $ & \textcolor{marker14}{$\blacksquare$}	\\
			$	5.0	$ & $	4.0	$ & $	100	$ & $	5.7	$ & $	1380	$ & $	78	$ & $	2.7e4	 $ & \textcolor{marker15}{$\blacksquare$}	\\
			$	4.9	$ & $	5.0	$ & $	100	$ & $	5.7	$ & $	1380	$ & $	60	$ & $	2.5e4	 $ & \textcolor{marker16}{$\blacksquare$}	\\
			$	5.0	$ & $	4.0	$ & $	128	$ & $	3.5	$ & $	1359	$ & $	61	$ & $	1.4e4 $ & \textcolor{marker17}{$\blacktriangledown$}	\\
			$	5.0	$ & $	4.0	$ & $	150	$ & $	4.9	$ & $	1257	$ & $	52	$ & $	1.0e4	 $ & \textcolor{marker18}{$\blacksquare$}	\\
			$	4.0	$ & $	4.0	$ & $	106	$ & $	4.1	$ & $	1331	$ & $	47	$ & $	6.7e3	 $ & \textcolor{marker19}{$\blacktriangledown$}	\\
			$	5.0	$ & $	4.0	$ & $	260	$ & $	4	$ & $	1400	$ & $	30	$ & $	2.0e3	 $ & \textcolor{marker20}{$\blacktriangledown$}	\\
			
		\end{tabular}
		\caption{Experimental parameters for figure 6 of the main text.}
		\label{TAB:Figure5}
	\end{center}
\end{table}

\begin{figure}[ht]
	\centering
	\includegraphics[width=0.65\linewidth]{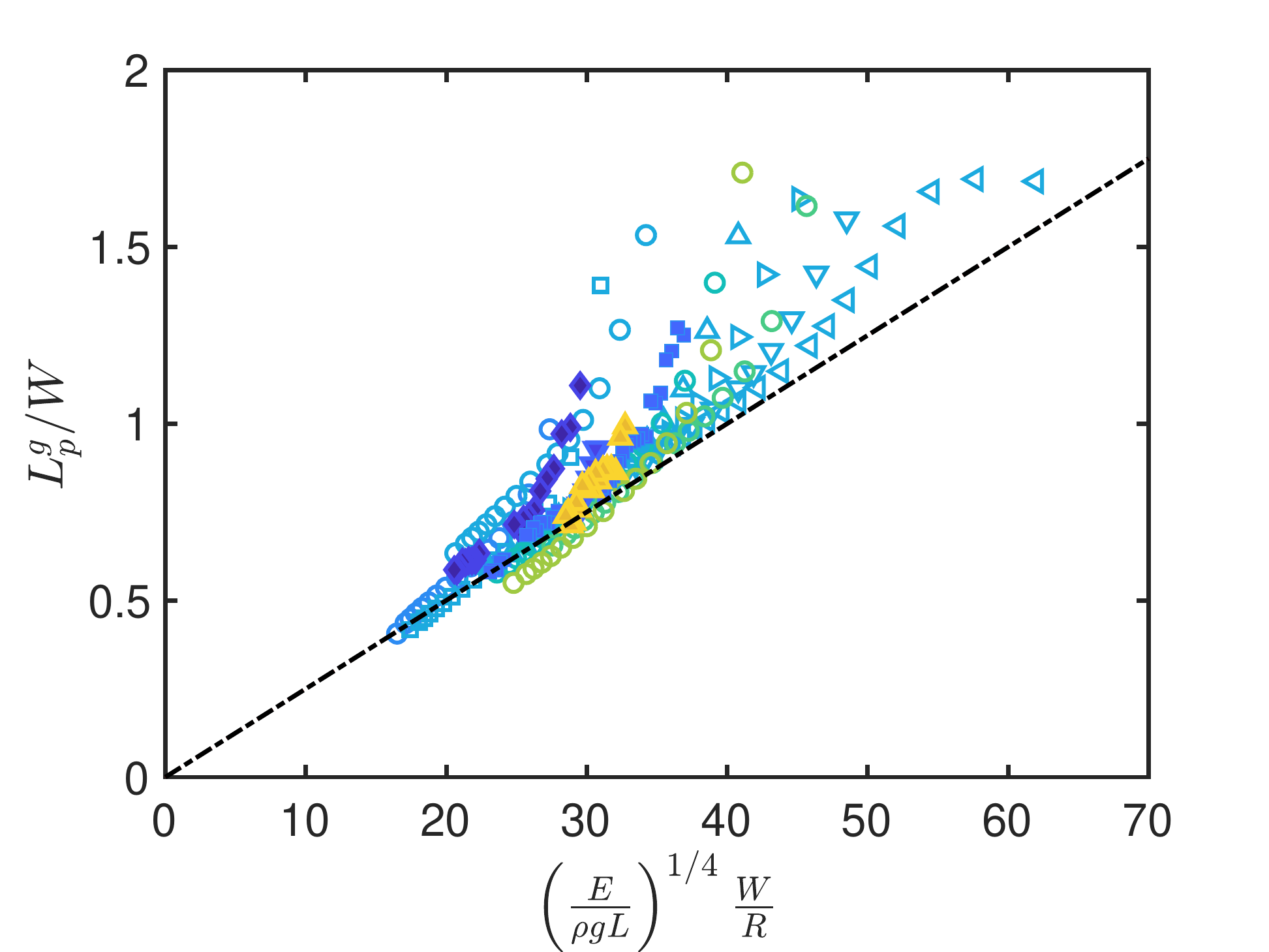}
	\caption{The persistence length $L_p^g$ well beyond the buckling threshold, measured for different geometrical and constitutive parameters (Table \ref{TAB:Figure7}) in experiments (closed symbols) and finite element simulations (open symbols), collapse onto a single trend upon normalization: the slope of the black dash-dotted line is derived as a best fit of the FEM data to be $0.025$}\label{FIG:LpG} 
\end{figure}

In table \ref{TAB:Figure7} we list the experimental parameter used to plot figure 8 in the main paper, reported here as figure \ref{FIG:LpG}, together with the marker shapes and colours used in this plot.

\definecolor{marker1}{RGB}{45,140,243}
\definecolor{marker2}{RGB}{28,170,223}
\definecolor{marker3}{RGB}{28,170,223}
\definecolor{marker4}{RGB}{28,170,223}
\definecolor{marker5}{RGB}{28,170,223}
\definecolor{marker6}{RGB}{28,170,223}
\definecolor{marker7}{RGB}{28,170,223}
\definecolor{marker8}{RGB}{18,190,185}
\definecolor{marker9}{RGB}{72,203,134}
\definecolor{marker10}{RGB}{159,201,66}
\definecolor{marker11}{RGB}{62,38,168}
\definecolor{marker12}{RGB}{71,67,231}
\definecolor{marker13}{RGB}{67,103,253}
\definecolor{marker14}{RGB}{234,186,48}

\begin{table}[H]
	\begin{center}
		\begin{tabular}{cccccccc}
			$W$ [cm]& $R$ [cm] & $t$ [$\mu$\,m] & $E$ [G\,Pa] & $\rho$ [kgm$^{-3}$] & $\kappa$ & Marker & \\
			$	5.0	$ & $	5.0	$ & $	300	$ & $	1	$ & $	900	$ & $	21	$ & \textcolor{marker1}{$\bigcirc$} & FEM \\
			$	5.0	$ & $	4.0	$ & $	300	$ & $	0.5	$ & $	900	$ & $	26	$ & \textcolor{marker2}{$\square$} & FEM		\\
			$	5.0	$ & $	4.0	$ & $	300	$ & $	1	$ & $	900	$ & $	26	$ & \textcolor{marker3}{$\bigcirc$} & FEM		\\
			$	5.0	$ & $	4.0	$ & $	300	$ & $	2	$ & $	900	$ & $	26	$ & \textcolor{marker4}{$\triangle$} & FEM		\\
			$	5.0	$ & $	4.0	$ & $	300	$ & $	3	$ & $	900	$ & $	26	$ & \textcolor{marker5}{$\rhd$} & FEM		\\
			$	5.0	$ & $	4.0	$ & $	300	$ & $	5	$ & $	900	$ & $	26	$ & \textcolor{marker6}{$\Diamond$} & FEM		\\
			$	5.0	$ & $	4.0	$ & $	300	$ & $	8	$ & $	900	$ & $	26	$ & \textcolor{marker7}{$\lhd$} & FEM		\\
			$	5.0	$ & $	3.5	$ & $	300	$ & $	1	$ & $	900	$ & $	30	$ & \textcolor{marker8}{$\bigcirc$} & FEM		\\
			$	5.0	$ & $	3.0	$ & $	300	$ & $	1	$ & $	900	$ & $	35	$ & \textcolor{marker9}{$\bigcirc$} & FEM		\\
			$	6.0	$ & $	4.0	$ & $	300	$ & $	1	$ & $	900	$ & $	38	$ & \textcolor{marker10}{$\bigcirc$} & FEM		\\
			$	4.0	$ & $	4.0	$ & $	395	$ & $	1.3	$ & $	898	$ & $	12.7	$ & \textcolor{marker11}{$\blacklozenge$} & EXP	\\
			$	5.0	$ & $	4.0	$ & $	484	$ & $	1.0	$ & $	905	$ & $	16.1	$ & \textcolor{marker12}{$\blacktriangledown$} & EXP	\\
			$	5.0	$ & $	4.0	$ & $	395	$ & $	1.3	$ & $	898	$ & $	19.8	$ & \textcolor{marker13}{$\blacksquare$} & EXP	\\
			$	5.0	$ & $	5.0	$ & $	128	$ & $	3.5	$ & $	1359	$ & $	48.8	$ & \textcolor{marker14}{$\blacktriangle$} & EXP	\\
			
		\end{tabular}
		\caption{Experimental parameters for figure 8 of the main text.}
		\label{TAB:Figure7}
	\end{center}
\end{table} 

\vskip2pc